\documentclass[10pt, journal, twocolumn]{IEEEtran}

\usepackage{amsmath}

\usepackage{amssymb}
\usepackage{bm}
\usepackage{color}
\usepackage{graphicx}
\usepackage{empheq}
\usepackage[linesnumbered,ruled,lined]{algorithm2e}
\usepackage[outdir=./]{epstopdf}
\usepackage{caption}
\usepackage{lipsum}
\usepackage{cite}
\usepackage{multirow}
\usepackage{subcaption}
\usepackage{fancyhdr}
\usepackage{graphicx}
\SetKwInput{Input}{input}
\SetKwInput{Output}{output}

\usepackage{array}
\newcolumntype{L}[1]{>{\raggedright\let\newline\\\arraybackslash\hspace{0pt}}m{#1}}
\newcolumntype{C}[1]{>{\centering\let\newline\\\arraybackslash\hspace{0pt}}m{#1}}
\newcolumntype{R}[1]{>{\raggedleft\let\newline\\\arraybackslash\hspace{0pt}}m{#1}}
\IEEEoverridecommandlockouts

\begin{document}

	\title{Rate-Splitting for Multi-Antenna Non-Orthogonal Unicast and Multicast Transmission: Spectral and Energy Efficiency Analysis
   }

		\author{
			 
		
			
			\IEEEauthorblockN{Yijie Mao,  Bruno Clerckx, \IEEEmembership{Senior Member, IEEE,
				} and Victor O.K. Li, \IEEEmembership{Fellow, IEEE
			} }	
			 \\\vspace{-4mm}
			 \thanks{Y. Mao and B. Clerckx are with Imperial College London, London SW7 2AZ, UK (email: y.mao16@imperial.ac.uk; b.clerckx@imperial.ac.uk). V.O.K. Li is with The University of Hong Kong, Hong Kong, China (email: vli@eee.hku.hk).
		This work has been partially supported by the U.K. Engineering and Physical
				Sciences Research Council (EPSRC) under grant EP/N015312/1, EP/R511547/1.
		A preliminary version of this paper was presented at the 19th IEEE international workshop on Signal Processing Advances in Wireless Communications (SPAWC) 2018\cite{mao2018rate}.}
				
			}

\maketitle

\thispagestyle{empty}
\pagestyle{empty}
\begin{abstract}
In a Non-Orthogonal Unicast and Multicast  (NOUM) transmission system, a multicast stream intended to all the receivers is superimposed in the power domain on the unicast streams. One layer of Successive Interference Cancellation (SIC) is required at each receiver to remove the multicast stream before decoding its intended unicast stream. In this paper, we first show that a linearly-precoded 1-layer Rate-Splitting (RS) strategy at the transmitter can efficiently exploit this existing SIC receiver architecture. By splitting the unicast messages into common and private parts and encoding the common parts along with the multicast message into a super-common stream decoded by all  users, the SIC is better reused for the dual purpose of separating the unicast and multicast streams as well as better managing the multi-user interference among the unicast streams. We further propose multi-layer  transmission strategies based on the generalized RS and power-domain Non-Orthogonal Multiple Access (NOMA).
Two different objectives are studied for the design of the precoders, namely, maximizing the Weighted Sum Rate (WSR) of the unicast messages and maximizing the system Energy Efficiency (EE), both subject to Quality of Service (QoS) rate requirements of all messages and a sum power constraint. A Weighted Minimum Mean Square Error (WMMSE)-based algorithm and a Successive Convex Approximation (SCA)-based algorithm are proposed  to solve the WSR and EE problems, respectively. Numerical results show that the proposed RS-assisted NOUM transmission strategies are more spectrally and energy efficient than the conventional Multi-User Linear-Precoding (MU--LP), { Orthogonal Multiple Access (OMA)} and power-domain NOMA in a wide range of user deployments (with a diversity of channel directions, channel strengths and qualities of channel state information at the transmitter) and network loads (underloaded and overloaded regimes).  It is superior for the downlink multi-antenna NOUM transmission.

\end{abstract}

\vspace{-3mm}
\section{Introduction}
\vspace{-1mm}

\par 
Two essential services, namely, unicast where each message is intended for a single user and multicast where each message is intended for multiple users,  are commonly supported in wireless networks.
Advanced wireless devices continue to strive for higher data rates of unicast services. Recently, the demands for multicast services, such as media streaming, mobile TV have been growing exponentially. 
Motivated by  the scarcity of the radio resources in the Fifth Generation (5G), researchers have focused on \textit{Non-Orthogonal Unicast and Multicast  (NOUM)} transmission \cite{unimulticast2008,unimulticast2010second,unimulti2012, Zhao2016LDM,Liu2017LDM,chen2017joint,tervo2017energy} where the unicast and multicast services are enabled in the same time-frequency resource blocks.  Such a transmission also finds applications as  Layered Division Multiplexing (LDM) in the digital TV standard ATSC 3.0 \cite{LDM2016} and recent interest for 5G in the 3rd Generation Partnership Project (3GPP) on concurrent delivery of both unicast and multicast services to  users and efficient multiplexing of multicast and unicast in time and frequency domains \cite{3gpp38913}. LDM has been shown to achieve a higher spectral efficiency than Time Division Multiplexing (TDM)/Frequency Division Multiplexing (FDM) in \cite{ldmtdm2015simeone}. 
From an information-theoretic perspective, Superposition Coding (SC) combined with Dirty Paper Coding (DPC) is first investigated in \cite{Weing2006Capacity} and further proved in \cite{Capacity2014Geng} to achieve the capacity region of the two-user NOUM transmission system. 

Due to the high computational burden of implementing DPC, Multi-User Linear Precoding (MU--LP) becomes the most attractive alternative to simplify the transmitter design. At the transmitter, the multicast stream intended for all users and the independent unicast streams are linearly precoded and superimposed before being sent to the users. At each user, the multicast stream is first decoded and removed  using Successive Interference Cancellation (SIC) and then  the intended unicast stream is decoded by fully treating any residual interference as noise. Such MU--LP-assisted NOUM has been studied previously with the objective of minimizing the transmit power \cite{Zhao2016LDM,Liu2017LDM},  maximizing the Weighted Sum Rate (WSR) \cite{chen2017joint} or the Energy Efficiency (EE) \cite{tervo2017energy}. The benefit of MU--LP-assisted transmission is to exploit all spatial multiplexing gains of a multi-antenna Broadcast Channel (BC) with perfect Channel State Information at the Transmitter (CSIT).
However, MU--LP is mainly suited to the underloaded regime (where the number of streams is smaller than the number of transmit antennas). It is sensitive to the user channel orthogonality and strengths, and does not optimally exploit the multiplexing gain of a multi-antenna BC with imperfect CSIT \cite{mao2017rate}. Moreover, the presence of SIC at the receivers is not exploited to manage the interference among the unicast streams, but only to separate the multicast stream from the unicast streams. In this paper, we resolve the above limitations of conventional MU--LP-assisted NOUM by resorting to linearly-precoded Rate-Splitting (RS) approaches.

\par Rate-Splitting was originally developed for the two-user single-antenna Interference Channel (IC) \cite{TeHan1981} and has recently been introduced  in \cite{RSintro16bruno} as a promising multi-user multi-antenna non-orthogonal transmission strategy to tackle numerous problems faced by modern Multiple Input Multiple Output (MIMO) wireless networks. Uniquely, RS enables to partially decode the interference and partially treat the interference as noise. This allows RS to explore a more general and powerful transmission framework, namely, Rate-Splitting Multiple Access (RSMA) for downlink multi-antenna systems that contains MU--LP and power-domain  Non-Orthogonal Multiple Access (NOMA) as special cases, and provides room for rate and Quality of Service (QoS) enhancements \cite{mao2017rate}. Though originally introduced for the two-user Single Input Single Output (SISO) IC, RS has recently appeared as an underpinning communication-theoretic strategy to tackle modern interference-related problems and has been successfully investigated in several multi-antenna broadcast channel settings, namely, unicast-only transmission with perfect CSIT \cite{ mao2017rate, mao2018EE, mao2018networkmimo,SYang2018SPAWC, Ahmad2018SPAWC}  and imperfect CSIT \cite{DoF2013SYang,RS2015bruno,RS2016joudeh, Minbo2016MassiveMIMO,RS2016hamdi, AG2016Gdof,  AP2017bruno,minbo2017mmWave, AG2017Gdof, enrico2017bruno,chenxi2017bruno,Lu2018MMSERS,Medra2018SPAWC,Flores2018ISWCS}, as well as (multigroup) multicast-only transmission \cite{hamdi2017bruno,Tervo2018SPAWC}.
With RS, each stream is split  at the transmitter into a common part and a private part. The common parts are jointly encoded into one common stream  to be decoded by all users while the private parts are independently encoded into  the private streams  to be decoded by the intended users. 
Upon decoding the common stream and the private stream, a user can reconstruct its original message. Due to the superimposed transmission of the common and private streams, RS can be viewed mathematically as a NOUM system. Hence, RS was termed joint multicasting and broadcasting in \cite{hamdi2015multicasting}.  
Though both the common stream in the RS-assisted transmission and the conventional multicast stream are decoded by multiple users, they are transmitted with different intentions. The multicast stream contains a single message intended for all those users  (because users are genuinely interested in the same message). On the other hand, the common stream in RS contains parts of the unicast messages of a subset of users, is intended to that subset of users, and is transmitted for interference management purposes. { All of the existing works on RS only considered unicast-only or multicast-only transmissions. The benefits of RS in NOUM transmissions have not been investigated yet.}

\par Motivated by the benefits of RS in the unicast-only and multicast-only transmissions as well as the limitations of conventional MU--LP-assisted NOUM, we study the application of RS in the NOUM transmission in this paper. The contributions of the paper are summarized as follows.

\textit{First}, we propose a 1-layer RS-assisted NOUM transmission strategy and design the precoder  to maximize WSR and EE, respectively. By splitting the unicast streams into common and private parts and encoding the common parts along with the multicast message into a super-common stream to be decoded by all users, the SIC in 1-layer RS is used for the dual purpose of separating the unicast and multicast streams as well as managing the interference among the unicast streams. The key benefit of 1-layer RS in the NOUM transmission is the fact that 1-layer RS does not lead to any complexity increase for the receivers compared to conventional MU--LP-assisted NOUM since one layer of SIC is required to separate multicast stream from unicast streams. This contrasts with unicast-only and muticast-only transmissions where 1-layer RS was found beneficial over MU--LP in \cite{RS2016hamdi,mao2017rate, mao2018EE} but at the cost of a receiver complexity increase due to the need of SIC for RS to operate. To the best of our knowledge, this is the first work that applies RS to  NOUM transmissions.
	
\textit{Second}, besides the 1-layer RS NOUM transmission strategy that incorporates a single layer of SIC, we further propose multi-layer SIC-assisted NOUM transmission strategies based on  the generalized RS and power-domain NOMA (referred to simply as NOMA in the rest of the paper).  NOMA relies on SC at the transmitter and SIC at the receivers (SC--SIC)\cite{NOMA2013YSaito}. It forces some users to fully decode and cancel the interference created by other users. Two NOMA-assisted NOUM transmission strategies are proposed, namely, `SC--SIC' and `SC--SIC per group'. To the best of our knowledge, this has not been investigated in the literature of multi-user multi-antenna NOUM transmissions.
Comparing with 1-layer RS, the proposed generalized RS allows the number of layers of the common streams to be increased with the number of served users. Thanks to its ability of partially decoding interference and partially treating interference as noise, the generalized RS model proposed in this work is a more general framework of multi-user multi-antenna NOUM transmission that encompasses MU--LP and NOMA as special cases.
	
\textit{Third}, we  study the WSR and EE maximization problems subject to the QoS rate requirements and a sum power constraint for all investigated NOUM strategies. Two optimization frameworks are proposed to solve the WSR and EE maximization problems based on the Weighted Minimum Mean Square Error (WMMSE)  and Successive Convex Approximation (SCA) algorithms, respectively. The effectiveness of the proposed algorithms is verified in the numerical results.

\textit{Fourth}, we show through numerical results that the proposed 1-layer RS-assisted NOUM transmission strategy is more spectrally and energy efficient than the existing MU--LP-assisted transmission in a wide range of user deployments (with a diversity of channel directions, channel strengths and qualities of channel state information at the transmitter) and network loads (underloaded and overloaded regimes). Importantly, applying 1-layer RS to NOUM boosts WSR and EE of the system but maintains the same receiver complexity as MU--LP. Hence, the performance gain comes at no additional cost for the receivers since one layer of SIC is required to separate unicast and multicast streams in the conventional MU--LP-assisted NOUM. In other words, 1-layer RS makes a better use of the existing SIC architecture. Comparing with the proposed NOMA-assisted NOUM, 1-layer RS achieves a more robust WSR and EE performance in a wide range of user deployments and network loads while its receiver complexity is much lower. 
	
\textit{Fifth}, we show that the WSR and EE performance of the proposed generalized RS is always equal to or larger than that of MU--LP and NOMA  { in the realm of NOUM transmissions.}  It is also more robust to the variation of user deployments, CSIT inaccuracy and network loads. As a consequence, the generalized RS is less sensitive to user pairing and therefore does not require complex user scheduling. The generalized RS requires a 
higher encoding and decoding complexity than MU--LP and NOMA since  multiple common streams are required to be encoded on top of the private streams.  { The observations in this paper confirm the superiority of RS over MU--LP, Orthogonal Multiple Access (OMA) where the unicast stream is only intended for a single user, and NOMA in NOUM transmissions, and complement our previous findings in \cite{mao2017rate,mao2018EE,hamdi2017bruno,RS2016hamdi} that have shown the superiority of RS in unicast-only and multicast-only transmissions. }


\par The rest of the paper is organized as follows. Section \ref{sec: system power model} introduces the system and power model. Section \ref{sec: one layer SIC} reviews the conventional MU--LP-assisted NOUM and the proposed 1-layer RS strategy. Section \ref{sec: multi layer SIC} specifies the proposed generalized RS and NOMA-assisted NOUM.  Section \ref{sec: algorithm} discusses the optimization frameworks to solve the WSR and EE  problems.  Section \ref{sec: WSR simulation} and \ref{sec: EE simulation} illustrate numerical results of WSR and EE. Section \ref{sec: conclusion} concludes the paper. 


\vspace{-2mm}
\section{System Model and Power Model}
\vspace{-0.5mm}
\label{sec: system power model}
\par Consider a BS equipped with $N_t$ antennas serving $K$ single-antenna users in the user set $\mathcal{K}=\{1,\ldots,K\}$. In each time frame, user-$k, \forall k\in\mathcal{K}$ requires a dedicated unicast message $W_k$ and a multicast message $W_0$. At the BS, the multicast message $W_0$  intended for all users and the $K$ unicast messages $W_1,\ldots,W_K$ are encoded into the data stream vector $\mathbf{s}$ and linearly precoded using the precoder $\mathbf{P}$. The transmit signal vector  $\mathbf{x}=\mathbf{P}\mathbf{s}$ is subject to the power constraint $\mathbb{E}\{||\mathbf{x}||^2\}\leq P_{t}$.   Assuming that $\mathbb{E}\{\mathbf{{s}}\mathbf{{s}}^H\}=\mathbf{I}$, we have $\mathrm{tr}(\mathbf{P}\mathbf{P}^{H})\leq P_{t}$. 
The signal received at user-$k$ is 
$
	y_{k}=\mathbf{{h}}_{k}^{H}\mathbf{{x}}+n_{k},
$
where $\mathbf{{h}}_{k}\in\mathbb{C}^{N_{t}\times1}$ is the channel between the BS and user-$k$, it is assumed to be perfectly known at the transmitter and receivers. The imperfect CSIT scenario will be discussed in the proposed algorithm and numerical results. The received noise $n_{k}$ is modeled as a complex Gaussian random variable with zero mean and variance $\sigma_{n,k}^{2}$. Without loss of generality, we assume the noise variances are equal to one ($\sigma_{n,k}^{2}=1,\forall k\in\mathcal{K}$). Hence, the transmit Signal-to-Noise Ratio (SNR) is equal to the transmit power consumption.
\par In this work, the total power consumption at the BS is \cite{xu2011improving}
\begin{equation}
P_{\textrm{total}}=\frac{1}{\eta}\mathrm{tr}\left(\mathbf{P}\mathbf{P}^{H}\right)+P_{\textrm{cir}},
\vspace{-1.5mm}
\end{equation}
where  $\eta\in[0,1]$ is the  power amplifier efficiency.  $P_{\textrm{cir}}=N_tP_{\textrm{dyn}}+P_{\textrm{sta}}$ is the circuit power consumption of the BS, where  $P_{\textrm{dyn}}$ is  the dynamic power consumption of one active  radio frequency chain and  $P_{\textrm{sta}}$ is the static power consumption of the cooling systems, power supply and so on. $\eta$ and $P_{\textrm{sta}}$ are assumed to be fixed for simplicity. 
\begin{figure}
	\centering
	\hspace*{0.2cm}
	\begin{subfigure}[b]{0.4\textwidth}
		\vspace{-0.1mm}
		\centering
		\includegraphics[width=0.8\textwidth]{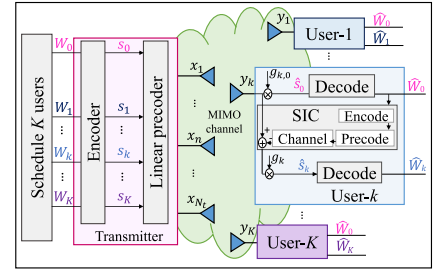}%
		\vspace{-0.1mm}
		\caption{MU--LP-assisted NOUM}
		\label{fig: mulp transmission model}
	\end{subfigure}%
	\\
	\hspace*{0.2cm}
	\begin{subfigure}[b]{0.4\textwidth}
		\vspace{2mm}
		\centering
		\includegraphics[width=\textwidth]{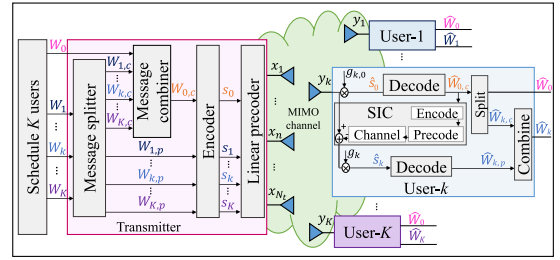}%
		\vspace{-1mm}
		\caption{1-layer RS-assisted NOUM}
		\label{fig: transmission model}
	\end{subfigure}%
	\vspace{-1mm}
	\caption{$K$-user one-layer SIC-based multi-antenna NOUM transmission model}
	\label{fig: 1-layer SIC transmission}
	\vspace{-4mm}
\end{figure}


\vspace{-2mm}
\section{One-layer SIC-based transmission}
\vspace{-1mm}
\label{sec: one layer SIC}
\par In this section, we focus on the  NOUM transmission model that only requires one layer of SIC at each receiver. We first introduce the baseline MU--LP-assisted strategy followed by the proposed 1-layer RS-assisted NOUM transmission model.
\vspace{-8mm}
\subsection{MU--LP}
\vspace{-1mm}
\label{sec: mulp}

\par The conventional MU--LP-assisted NOUM transmission model is illustrated in Fig. \ref{fig: mulp transmission model}. The multicast message $W_0$ and the unicast messages $W_1,\ldots,W_K$ are independently encoded into the data streams $s_0,s_1,\ldots,s_K$.  The stream vector $\mathbf{s}=[s_0,s_1,\ldots,s_K]^{T}$ is precoded using the precoder $\mathbf{P}=[\mathbf{p}_0,\mathbf{p}_1,\ldots,\mathbf{p}_K]$, where $\mathbf{p}_0,\mathbf{p}_k\in\mathbb{C}^{N_t\times1}$ are the respective precoders of the multicast stream $s_0$ and the unicast stream $s_k$. The resulting transmit signal $\mathbf{x}\in\mathbb{C}^{N_t\times1}$  is
\vspace{-1mm}
\begin{equation}
\label{eq: mulp transmit signal}
\mathbf{x}=\mathbf{P}\mathbf{{s}}=
\underbrace{\mathbf{p}_{0}s_{0}}_{{\text{multicast stream}}}
+
\underbrace{\sum_{k\in\mathcal{K}}\mathbf{p}_{k}s_{k}}_{{\text{unicast streams}}}
.
\vspace{-2mm}
\end{equation}

\par The signal received at user-$k$  becomes
\begin{equation}
\label{eq: mulp receive signal}
\begin{aligned}
y_{k}
&=\underbrace{\mathbf{{h}}_{k}^{H}\mathbf{p}_{0}s_{0}}_{{\text{intended multicast stream}}}
+
\underbrace{\mathbf{{h}}_{k}^{H}\mathbf{p}_{k}s_{k}}_{{\text{intended unicast stream}}}
\\&+
\underbrace{\sum_{j\in\mathcal{K},j\neq k}\mathbf{{h}}_{k}^{H}\mathbf{p}_{j}s_{j}}_{{\text{interference among unicast streams}}}
+
\underbrace{n_{k}}_{{\text{noise}}}.
\end{aligned}
\end{equation}

\par Each user-$k,\forall k\in\mathcal{K}$ decodes the multicast stream $s_0$ and the intended unicast stream $s_k$ under the assistance of one SIC. The decoding order of $s_0$ and $s_k$ can be optimized for each instantaneous channel condition. 
The decoding order follows the rule that \textit{the data stream intended for more users has a higher decoding priority} \cite{Zhao2016LDM,chen2017joint}. 
Hence, we assume that the multicast stream is decoded first  and removed from the received signal using SIC before decoding the unicast streams at all users. This assumption will be applied to all the transmission strategies proposed in the rest of the paper.  The multicast stream $s_0$ is decoded by treating the signal of all unicast streams as noise. The Signal-to-Interference-plus-Noise Ratio (SINR) of decoding $s_0$ at user-$k$ is
\vspace{-0.8mm}
\begin{equation}
\label{eq: mulp common sinr}
\gamma_{k,0}=\frac{|\mathbf{{h}}_{k}^{H}\mathbf{{p}}_{0}|^{2}}{\sum_{j\in\mathcal{K}}|\mathbf{{h}}_{k}^{H}\mathbf{{p}}_{j}|^{2}+1}.
\vspace{-0.2mm}
\end{equation}
Once  $s_{0}$ is successfully decoded and subtracted from the original received signal  $y_k$, user-$k$ decodes the intended unicast stream $s_{k}$ by treating the interference from the unicast streams of other users as noise. The SINR of decoding $s_{k}$ at user-$k$ is
\vspace{-1mm}
\begin{equation}
\label{eq: mulp private sinr}
\gamma_{k}=\frac{|\mathbf{{h}}_{k}^{H}\mathbf{{p}}_{k}|^{2}}{\sum_{j\in\mathcal{K},j\neq k}|\mathbf{{h}}_{k}^{H}\mathbf{{p}}_{j}|^{2}+1}.
\vspace{-0.2mm}
\end{equation}
The corresponding achievable rates of decoding $s_{0}$ and $s_{k}$ at user-$k$ are
$R_{k,0}=\log_{2}\left(1+\gamma_{k,0}\right)$, 
$R_{k}=\log_{2}\left(1+\gamma_{k}\right)$. 
As  $s_0$ is decoded by all users, to ensure that $s_{0}$ is successfully decoded by all users, the corresponding code-rate should not exceed the rate achievable by the weakest receiver   \cite{RS2016joudeh,hamdi2017bruno}, which is given by
\vspace{-2mm}
\begin{equation}
\label{eq: mulp multicast rate}
R_{0}=\min\left\{ R_{1,0},\ldots,R_{K,0}\right\}.
\vspace{-2mm}
\end{equation} 
%

\par Two different objectives are studied for the design of the precoders:

\par \textit{1) Weighted sum rate maximization problem:} To investigate the spectral efficiency, we study the problem of maximizing the WSR of the unicast messages while the QoS rate constraints of all messages and the power constraint of the BS should be met. For a given weight vector $\mathbf{u}=[u_1,\ldots,u_K]$, the WSR maximization problem in the $K$-user MU--LP-assisted NOUM is
	\vspace{-3mm}
\begin{subequations}
	\label{eq: mulp}
	\begin{empheq}[left=\textrm{WSR}_{\textrm{MU--LP}}\empheqlbrace]{align}
	\max_{\mathbf{{P}}}\,\,&\sum_{k\in\mathcal{K}}u_{k}R_{k}  \\
	\mbox{s.t.}\quad
	&  \,\,R_{k}\geq R_{k}^{th}, \forall k\in\mathcal{K},  \label{c1_mulp}\\
	&\,\, R_{k,0}\geq R_0^{th},\forall k\in\mathcal{K},   \label{c2_mulp}\\
	&\,\,	\text{tr}(\mathbf{P}\mathbf{P}^{H})\leq P_{t},   \label{c3_mulp}	
	\end{empheq}
\end{subequations}	
where Constraint (\ref{c1_mulp}) is the QoS rate requirement of each unicast message. $R_k^{th}$ is the rate lower bound of the unicast message $W_k$.  Constraint (\ref{c2_mulp}) ensures that each user decodes the multicast message $W_0$ with a rate larger than or equal to $R_0^{th}$. 

\textit{2) Energy efficiency maximization problem:} To investigate the EE of MU--LP, we maximize the WSR of all the messages divided by the sum power of the transmitter. For a given weight vector $\mathbf{u}_{tot}=[u_0,u_1,\ldots,u_K]$ of all the messages, the EE maximization problem of MU--LP is
 \begin{equation}
	\label{eq: EE MULP}
	\textrm{EE}_{\textrm{MU--LP}}\begin{dcases}
		\max_{\mathbf{{P}}}\,\,&\frac{u_{0}R_{0}+\sum\limits_{k\in\mathcal{K}}u_{k}R_{k} }{\frac{1}{\eta}\mathrm{tr}(\mathbf{P}\mathbf{P}^{H})+P_{\textrm{cir}}}  \\
		\mbox{s.t.}\quad
		& \textrm{(\ref{c1_mulp})--(\ref{c3_mulp})}.
	\end{dcases}
\end{equation}

\par \textit{Remark 1: Recall that MU--LP does not require any SIC at each user in the unicast-only transmission. In comparison, one layer of SIC is necessary at each user to remove the multicast stream before decoding the
intended unicast stream in the MU--LP-assisted NOUM transmission. The SIC is used for the purpose of separating the unicast and multicast streams.}
 
\vspace{-4mm}
 \subsection{1-layer RS}
 \vspace{-1mm}
\label{sec: 1-layer RS}

\par The proposed $K$-user 1-layer RS-assisted NOUM transmission model is illustrated in Fig. \ref{fig: transmission model}. 
The unicast message $W_k$ intended for user-$k$, $\forall k\in\mathcal{K}$ is split into a common sub-message $W_{k,c}$ and a private sub-message $W_{k,p}$. The private sub-messages  $W_{1,p},\ldots,W_{K,p}$ of the unicast messages are independently encoded into the private streams  ${s}_1, \ldots,{s}_K$ while the common sub-messages $W_{1,c},\ldots,W_{K,c}$ of the unicast messages are jointly encoded with the multicast message $W_0$ into a super-common stream ${s}_0$ required to be decoded by all users. { Different from the common stream ${s}_0$ in MU--LP that only includes the multicast meesage, the super-common stream ${s}_0$ in 1-layer RS includes the whole multicast message as well as parts of the unicast messages. Following the transmission procedure in MU--LP,  the formed stream vector $\mathbf{s}$
 is linearly precoded and broadcast to the users.}	


\par The super-common stream and private streams are decoded using one layer of SIC in a similar way as decoding the multicast stream and the unicast streams in the MU--LP-assisted NOUM transmission with higher decoding priority given to the super-common stream.
Since $R_{0}$ is now shared by the achievable rates of transmitting the multicast message $W_0$ and the common sub-messages $W_{1,c},\ldots,W_{K,c}$  of the unicast messages, it is equal to
$
C_{0}+\sum_{k\in\mathcal{K}}C_{k,0}=R_{0}, 
$
where $C_0$  is the portion of $R_{0}$ transmitting $W_{0}$ and  $C_{k,0}$ is the user-$k$'s portion of $R_{0}$ transmitting $W_{k,c}$. The portions of rate allocated to $W_{0}$  and $W_{1,c},\ldots,W_{K,c}$ will be optimized by solving the optimization problems formulated in this section. In the proposed 1-layer RS-assisted NOUM transmission, the achievable rate of each unicast message contains two parts. One part is $C_{k,0}$ transmitted via $W_{k,c}$ encoded in the super-common stream $s_0$. The other part is $R_k$ transmitted via $W_{k,p}$ encoded in the private stream $s_k$. 
Hence, the achievable rate of transmitting the unicast message $W_k$ of user-$k$ is 
$
R_{k,tot}=C_{k,0}+R_{k}, \forall k\in\mathcal{K}.
$
The corresponding WSR and EE maximization problems are given by

\par  \textit{1) Weighted sum rate maximization problem:}
The WSR maximization problem in the $K$-user 1-layer RS-assisted NOUM transmission for a given $\mathbf{u}$ is
\vspace{-1mm}
\begin{subequations}
	\label{eq: rs}
	\begin{empheq}[left=\textrm{WSR}_{\textrm{1-layer RS}}\empheqlbrace]{align}\max_{\mathbf{{P}}, \mathbf{c}}\,\,&\sum_{k\in\mathcal{K}}u_{k}R_{k,tot}  \label{o1_rs}\\
	\mbox{s.t.}\quad
	&  \,\,C_{k,0}+R_{k}\geq R_{k}^{th}, \forall k \in \mathcal{K} \label{c1_rs}\\
	&\,\, C_{0}\geq R_0^{th}\label{c2_rs}\\
	&\,\, C_{0}+\sum_{j\in\mathcal{K}}C_{j,0}\leq R_{k,0},\forall k\in\mathcal{K} \label{c3_rs}\\
	&\,\, C_{k,0}\geq 0,\forall k\in\mathcal{K} \label{c4_rs}\\
	&\,\, \text{tr}(\mathbf{P}\mathbf{P}^{H})\leq P_{t} \label{c5_rs}
	\end{empheq}
\end{subequations}
where  $\mathbf{c}=[C_0,C_{1,0},\ldots,C_{K,0}]$ is the common rate vector required to be optimized with the precoder $\mathbf{P}.$
When $C_{k,0}=0, \forall k\in\mathcal{K}$, Problem $\textrm{WSR}_{\textrm{1-layer RS}}$ reduces to Problem $\textrm{WSR}_{\textrm{MU--LP}}$. Hence, the proposed RS model always achieves the same or superior performance to MU--LP. Constraint (\ref{c3_rs}) ensures the super-common stream can be successfully decoded by all users. Constraints (\ref{c1_rs}) and (\ref{c2_rs}) are the QoS rate constraints of all  messages.

\par \textit{2) Energy efficiency maximization problem:}
The EE maximization problem of 1-layer RS for a given $\mathbf{u}_{tot}$ is
\vspace{-1mm}
\begin{equation}
\label{eq: EE tot RSMA}
\textrm{EE}_{\textrm{1-layer RS}}\begin{dcases}
\max_{\mathbf{c},\mathbf{{P}}}\,\,&\frac{u_{0}C_{0}+\sum_{k\in\mathcal{K}}u_kR_{k,tot}}{\frac{1}{\eta}\mathrm{tr}(\mathbf{P}\mathbf{P}^{H})+P_{\textrm{cir}}}  \\
\mbox{s.t.}\quad
& \textrm{(\ref{c1_rs})--(\ref{c5_rs})}.
\end{dcases}
\end{equation}

\par \textit{Remark 2: Similarly to the $K$-user 1-layer RS-assisted unicast-only transmission discussed in \cite{mao2017rate}, one layer of SIC is required at each user in the $K$-user 1-layer RS-assisted NOUM transmission. In contrast with the MU--LP-assisted NOUM, the SIC of 1-layer RS-assisted NOUM transmission is used for separating the unicast and multicast streams as well as better managing the multi-user interference among the unicast streams. The
presence of SIC is therefore better exploited in the 1-layer RS-assisted NOUM than in the MU--LP-assisted NOUM.}


\vspace{-2mm}
\section{Multi-layer SIC-based transmission}
\label{sec: multi layer SIC}
To further enhance the  system spectral and energy efficiencies, the co-channel interference among unicast streams can be better managed by introducing multiple layers of SIC at each receiver to decode part of the interference. There are two multi-layer SIC-based transmission strategies, namely, RSMA and NOMA-based transmission. In the unicast-only transmission, it has been shown in \cite{mao2017rate,mao2018EE} that NOMA achieves better spectral and energy efficiency than MU--LP when the user channels are aligned and there is certain channel strength difference among users. 
The generalized RS-based RSMA bridges MU--LP and NOMA and achieves a better spectrum efficiency \cite{mao2017rate}. In this section, both RSMA and NOMA strategies are applied to the NOUM transmission. To simplify the explanation, we focus on the three-user case ($\mathcal{K}=\{1,2,3\}$) for all multi-layer SIC transmission strategies. {It can be extended to solve the $K$-user problem.}

\vspace{-4mm}
\subsection{Generalized rate-splitting}
\vspace{-1mm}
\begin{figure}[t!]
	\vspace{-0mm}
	\centering
	\includegraphics[width=3.4in]{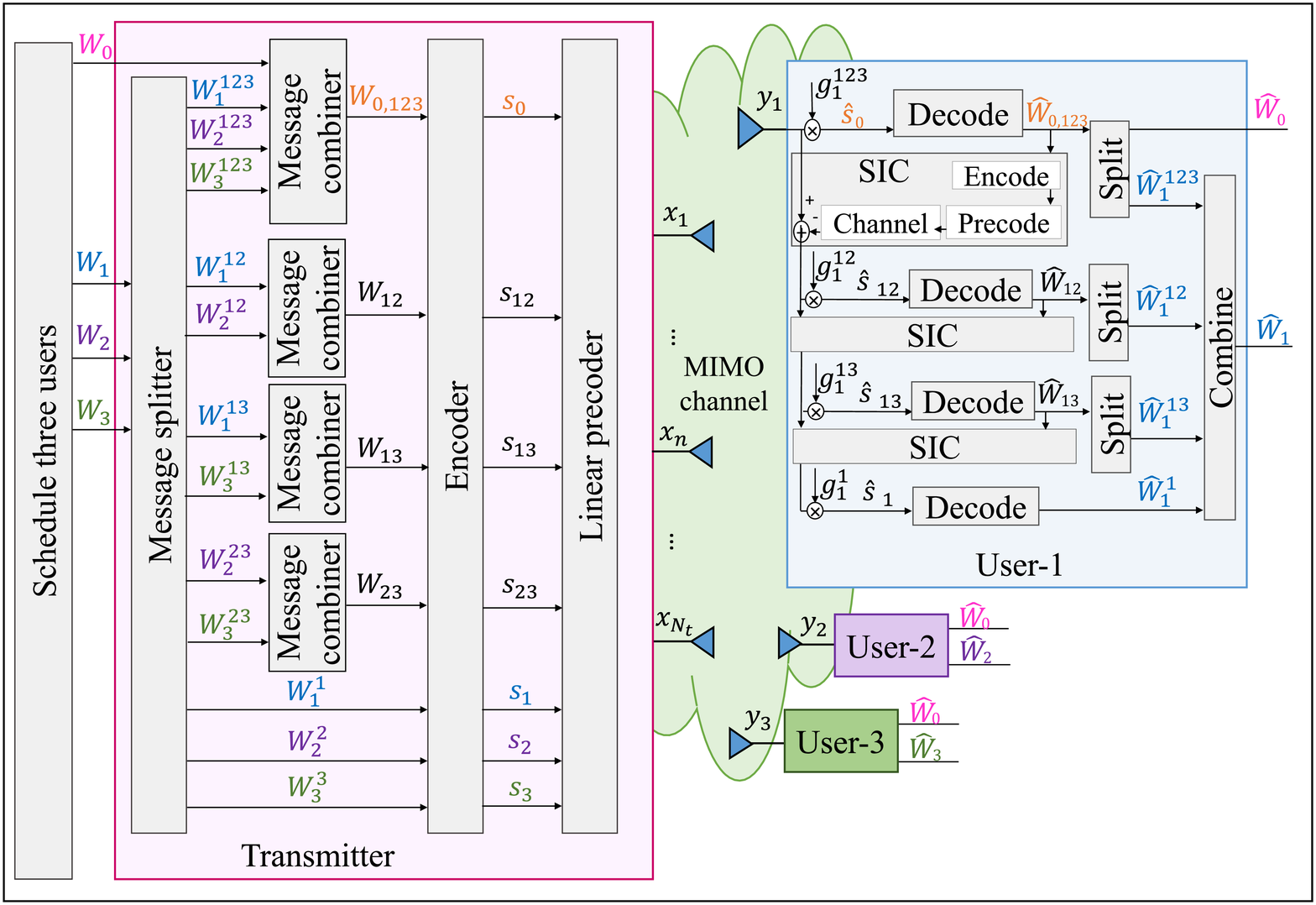}%
	\vspace{-0.2mm}
	\caption{Three-user generalized RS-assisted multi-antenna NOUM transmission model}
	\label{fig: transmission model generalized RS}
	\vspace{-5mm}
\end{figure}
\par  Different from the 1-layer RS transmission model introduced in Section \ref{sec: 1-layer RS} where the unicast message of each user is split into two parts, the unicast message of each user is split into four different parts in the three-user generalized RS transmission model. For user-$1$, the unicast message $W_1$ is split into sub-messages $\{W_{1}^{123}$, $W_{1}^{12}$, $W_{1}^{13}$, $W_{1}^{1}\}$. The unicast messages of user-$2$ and user-$3$ are split into sub-messages $\{W_{2}^{123}, W_{2}^{12},W_{2}^{23},W_{2}^{2}\}$ and  $\{W_{3}^{123}, W_{3}^{13},W_{3}^{23},W_{3}^{3}\}$, respectively. The superscript of each sub-message represents a group of users. The sub-messages with the same superscript are encoded together into a common stream intended for the users within that specific user group. Sub-messages $W_{1}^{123},W_{2}^{123},W_{3}^{123}$ are jointly encoded with the multicast message $W_0$ into the super-common stream $s_{0}$ intended for all the three users. Sub-messages $W_{1}^{12}, W_{2}^{12}$ are encoded together into the partial-common stream $s_{12}$ intended for user-$1$ and user-$2$ only. Similary, we obtain the partial-common streams $s_{13}$ and $s_{23}$ encoded by $W_{1}^{13}, W_{3}^{13}$ and $W_{2}^{23}, W_{3}^{23}$, respectively. Sub-messages $W_1^1, W_2^2, W_3^3$ are respectively encoded into the private streams $s_1, s_2,s_3$ for a single user only. The intention of splitting each unicast message into different sub-messages and reuniting the sub-messages is to enable each user the capability of dynamic interference management. For example, when user-$1$ decodes $s_{0}$, it not only decodes the intended multicast message $W_{0}$ and the intended unicast sub-message $W_{1}^{123}$ but also partially decodes the interference resulting from sub-messages $W_{2}^{123}$ and $W_{3}^{123}$. The encoded data streams $\mathbf{{s}}=[s_{0},s_{12},s_{13},s_{23},s_{1},s_{2},s_{3}]^T$ are precoded via the precoder $\mathbf{{P}}=[\mathbf{{p}}_{0},\mathbf{{p}}_{12},\mathbf{{p}}_{13},\mathbf{{p}}_{23},\mathbf{{p}}_{1},\mathbf{{p}}_{2},\mathbf{{p}}_{3}]$ and then broadcast to the users. The transmit signal $\mathbf{x}\in\mathbb{C}^{N_t\times1}$  is 
\vspace{-1mm}
\begin{equation}
\mathbf{x}=
\underbrace{\mathbf{p}_{0}s_{0}}_{\text{super-common stream}}+\underbrace{\sum_{i\in\{12,13,23\}}\mathbf{p}_{i}s_{i}}_{\text{partial-common streams}}+
\underbrace{\sum_{k\in\mathcal{K}}\mathbf{p}_{k}s_{k}}_{\text{private streams}} .
\vspace{-0.2mm}
\end{equation} 
\par At user sides, each user decodes the data streams that carry its intended sub-messages using SIC. 
The decoding procedure starts from the super-common stream to the partial-common streams and then progresses downwards to the private streams. 
At user-$1$, the data streams $s_{0}, s_{12}, s_{13}, s_{1}$ are decoded using SIC. Similarly, user-$2$ and user-$3$ decode the data streams $s_{0}, s_{12}, s_{23}, s_{2}$ and $s_{0}, s_{13}, s_{23}, s_{3}$, respectively. As $s_{12},s_{13},s_{23}$ are all intended for two users, the decoding order needs to be optimized together with the precoder $\mathbf{P}$. The decoding order of all streams intended for two users is denoted by $\pi_2$. For instance, when the decoding order is $\pi_2=12\rightarrow13\rightarrow23$,  $s_{12}$ will  be decoded before $s_{13}$ and $s_{13}$ will be decoded before $s_{23}$ at all users. Since user-1 only decodes the partial-common streams $s_{12}$ and $s_{13}$, the corresponding decoding order at user-$1$ is denoted by $\pi_{2,1}=12\rightarrow13$. We further use $s_{\pi_{2,k}(i)}$ to represent the $i$th data stream to be decoded at user-$k$ based on the decoding order $\pi_2$. When the decoding order at user-$1$ is $\pi_{2,1}=12\rightarrow13$, we have $s_{\pi_{2,1}(1)}=s_{12}$ and  $s_{\pi_{2,1}(2)}=s_{13}$. 
The proposed three-user generalized RS-assisted NOUM transmission model with the decoding order $\pi_2=12\rightarrow13\rightarrow23$ is illustrated in Fig. \ref{fig: transmission model generalized RS}.
The SINRs of decoding the streams $s_{0}, s_{\pi_{2,1}(1)}, s_{\pi_{2,1}(2)}, s_{1}$ using SIC  at user-$1$ are respectively given by
\vspace{-1mm}
\begin{equation}
\gamma_{1}^{0}=\frac{\left|\mathbf{{h}}_{1}^{H}\mathbf{{p}}_{0}\right|^{2}}{\sum_{i\in\{12,13,23\}}\left|\mathbf{{h}}_{1}^{H}\mathbf{{p}}_{i}\right|^{2}+\sum_{k=1}^3\left|\mathbf{{h}}_{1}^{H}\mathbf{{p}}_{k}\right|^{2}+1},
\end{equation}
\vspace{-1mm}
\begin{equation}
\gamma_{1}^{\pi_{2,1}{(1)}}=\frac{\left|\mathbf{{h}}_{1}^{H}\mathbf{{p}}_{\pi_{2,1}{(1)}}\right|^{2}}{\left|\mathbf{{h}}_{1}^{H}\mathbf{{p}}_{\pi_{2,1}{(2)}}\right|^{2}+\left|\mathbf{{h}}_{1}^{H}\mathbf{{p}}_{23}\right|^{2}+\sum_{k=1}^3\left|\mathbf{{h}}_{1}^{H}\mathbf{{p}}_{k}\right|^{2}+1},
\end{equation}
\vspace{-3.5mm}
\begin{equation}
\gamma_{1}^{\pi_{2,1}{(2)}}=\frac{\left|\mathbf{{h}}_{1}^{H}\mathbf{{p}}_{\pi_{2,1}{(2)}}\right|^{2}}{\left|\mathbf{{h}}_{1}^{H}\mathbf{{p}}_{23}\right|^{2}+\sum_{k=1}^3\left|\mathbf{{h}}_{1}^{H}\mathbf{{p}}_{k}\right|^{2}+1},
\end{equation}
\vspace{-1mm}
\begin{equation}
\gamma_{1}=\frac{\left|\mathbf{{h}}_{1}^{H}\mathbf{{p}}_1\right|^{2}}{\left|\mathbf{{h}}_{1}^{H}\mathbf{{p}}_{23}\right|^{2}+\sum_{k=2}^3\left|\mathbf{{h}}_{1}^{H}\mathbf{{p}}_{k}\right|^{2}+1}.
\end{equation}
The resulting achievable rates of decoding the intended streams at user-$1$ are calculated by $R_{1}^{i}=\log_{2}\left(1+\gamma_{1}^{i}\right), \forall i\in\{0,12,13,1\}$.
By using the same method, we could obtain the individual rates of decoding the intended streams at user-$2$ and user-$3$, respectively. 
To ensure that the streams are decodable by the corresponding groups of users, the transmission common rates should not exceed $R_{0}=\min\left\{ R_{1}^{0},R_{2}^{0},R_{3}^{0}\right\},R_{12}=\min\left\{ R_{1}^{12},R_{2}^{12}\right\},R_{13}=\min\left\{ R_{1}^{13},R_{3}^{13}\right\},R_{23}=\min\left\{ R_{2}^{23},R_{3}^{23}\right\}$.
Following the above RS structure, the rate of each common stream is split for the corresponding groups of users. Let $C_0$ be the portion of $R_0$ transmitting $W_0$ and $C_k^i$ be the portions of rate $R_i$  allocated to user-$k$ for the transmission of the sub-message $W_{k}^i$, we have { $C_0+\sum_{k\in\{1,2,3\}}C_{k}^{123}=R_{0}$, $\sum_{k\in\{1,2\}}C_k^{12}=R_{12}$, $\sum_{k\in\{1,3\}}C_k^{13}=R_{13}$, $\sum_{k\in\{2,3\}}C_k^{23}=R_{23}$.}
Hence, the individual  rate of transmitting the unicast message of each user is the summation of the portions of rate in the intended common streams, which is given by  {$R_{k,tot}=\sum_{i_k} C_{k}^{i_k}+R_{k}$, where $i_1\in\{0,12,13\}$, $i_2\in\{0,12,23\}$ and $i_3\in\{0,13,23\}$. }

\par The corresponding WSR and EE maximization problems are given by
\par  \textit{1) Weighted sum rate maximization problem:}
The WSR maximization problem in the three-user generalized RS-assisted NOUM transmission for a given $\mathbf{u}$ is
\vspace{-2mm}
\begin{subequations}
	\label{eq: general rs}
	\begin{empheq}[left=\textrm{WSR}_{\textrm{general RS}}\empheqlbrace]{align}\max_{\mathbf{{P}}, \mathbf{c},\pi}\,\,&\sum_{k\in\mathcal{K}}u_{k}R_{k,tot}  \label{o1_grs}\\
	\mbox{s.t.}\quad
	&  \,\,R_{k,tot}\geq R_k^{th},  \forall k\in\mathcal{K} \label{c1_grs}\\
	&\,\, C_{0}\geq R_0^{th}\label{c2_grs}\\
	&\,\, C_{0}+\sum_{j\in\mathcal{K}}C_j^{123}\leq R_{k}^0,\forall k\in\mathcal{K} \label{c3_grs}\\
	&\,\, C_1^{12}+C_2^{12} \leq R_{k}^{12},\forall k\in\{1,2\} \label{c4_grs}\\
	&\,\, C_1^{13}+C_3^{13} \leq R_{k}^{13},\forall k\in\{1,3\} \label{c5_grs}\\
	&\,\, C_2^{23}+C_3^{23} \leq R_{k}^{23},\forall k\in\{2,3\} \label{c6_grs}\\
	&\,\,  \mathbf{c}\geq \mathbf{0} \label{c7_grs}\\
	&\,\, \text{tr}(\mathbf{P}\mathbf{P}^{H})\leq P_{t} \label{c8_grs}
	\end{empheq}
\end{subequations}
where  $\mathbf{c}=[C_0, C_1^{123},C_2^{123},C_3^{123},C_1^{12},C_2^{12},C_1^{13},C_3^{13},C_2^{23},$ $C_3^{23}]$ is the common rate vector.
When there is zero common rate allocated to the sub-messages intended for two users, i.e., $C_k^{i}=0,\forall i\in\{12,13,23\}, k\in\mathcal{K}$, Problem $\textrm{WSR}_{\textrm{general RS}}$ reduces to Problem $\textrm{WSR}_{\textrm{1-layer RS}}$. When $C_k^{i}=0,\forall i\in\{123,12,13,23\}, k\in\mathcal{K}$, Problem $\textrm{WSR}_{\textrm{general RS}}$ reduces to $\textrm{WSR}_{\textrm{MU--LP}}$. Hence, the proposed generalized RS model always achieves the same or superior performance to 1-layer RS and MU--LP. Constraints (\ref{c3_grs})--(\ref{c6_grs}) ensures all common streams are decodable by the intended users. Constraints (\ref{c1_grs}) and (\ref{c2_grs}) are the QoS rate constraints.

\par \textit{2) Energy efficiency maximization problem:}
The EE maximization problem of the generalized RS for a given $\mathbf{u}_{tot}$ is
\vspace{-1mm}
\begin{equation}
\label{eq: EE tot gRSMA}
\textrm{EE}_{\textrm{general RS}}\begin{dcases}
\max_{\mathbf{{P}}, \mathbf{c},\pi}\,\,&\frac{u_{0}C_{0}+\sum_{k\in\mathcal{K}}u_kR_{k,tot}}{\frac{1}{\eta}\mathrm{tr}(\mathbf{P}\mathbf{P}^{H})+P_{\textrm{cir}}}  \\
\mbox{s.t.}\quad
& \textrm{(\ref{c1_grs})--(\ref{c8_grs})}.
\end{dcases}
\vspace{-1mm}
\end{equation}

\par \textit{Remark 3: The proposed generalized RS-based NOUM is a super-strategy of the 1-layer RS-based NOUM  proposed in Section \ref{sec: 1-layer RS}. 
{As more layers of SIC are required at each user to decode the partial-common streams},  the receiver complexity of the proposed generalized RS-based NOUM increases with the number of served users $K$. In comparison,  the receiver complexity of 1-layer RS does not depend on  $K$ and  is much lower especially when  $K$ is large.}

\vspace{-2mm}
\subsection{NOMA}
\label{sec: NOMA}
 \par
\begin{figure}
	\centering 
	\begin{subfigure}[b]{0.4\textwidth}	
		\centering
		\includegraphics[width=0.86\textwidth]{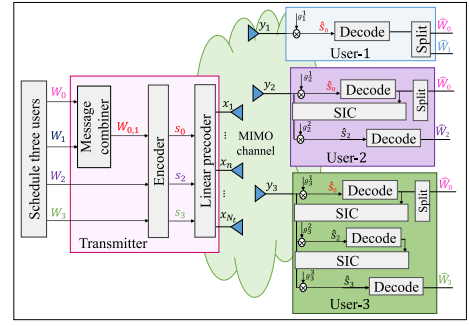}%
		\vspace{-1mm}
		\caption{SC--SIC-assisted NOUM}
		\vspace{2mm}
		\label{fig: scsic transmission model}
	\end{subfigure}%
\\
	\begin{subfigure}[b]{0.4\textwidth}
		\centering
		\includegraphics[width=\textwidth]{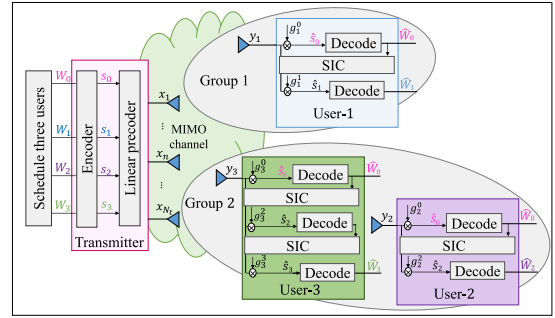}%
		\vspace{-0.5mm}
		\caption{SC--SIC per group-assisted NOUM}
		\label{fig: scsic per group transmission model}
	\end{subfigure}%
     \vspace{-0.5mm}
	\caption{Three-user NOMA-assisted multi-antenna NOUM transmission model}
	\label{fig: NOMA transmission}
	\vspace{-4mm}
\end{figure}

 There are two main strategies in the multi-antenna NOMA, namely, `SC--SIC' and `SC--SIC per group' \cite{mao2017rate}. Both are applied in the  NOUM transmission. 
 Comparing with the SC--SIC-assisted unicast-only transmission, the main difference in the SC--SIC-assisted NOUM transmission is that the multicast message $W_0$ is jointly encoded with the unicast message to be decoded first into a common stream $s_0$. At user sides, each user first decodes $s_0$ with the highest priority. Then the users carry on decoding the unicast streams according to the decoding order $\pi$. The proposed three-user SC--SIC-assisted NOUM transmission model with the decoding order $\pi=1\rightarrow2\rightarrow3$ is illustrated in Fig. \ref{fig: scsic transmission model}.  The first layer of SIC is used for two different purposes. It is used not only  to decode the multi-user interference among  the unicast streams, but also to separate the unicast and multicast streams.  The decoding order is required to be optimized with the precoder for  both WSR and EE optimization problem. 

 \begin{table*}[t!]
	\centering
	\caption{{Qualitative comparison of the complexity of different strategies for NOUM}}
	\label{tab: complexity}
	{	\begin{tabular}{|L{1.5cm}|L{2.2cm}|L{2.2cm}|L{2.4cm}|L{2.7cm}|L{2.5cm}|}
			\hline
			\textbf{Category}                 & \multicolumn{2}{c|}{ \textbf{One-layer SIC-based transmission}} & \multicolumn{3}{c|}{\textbf{Multi-layer SIC-based transmission}}       \\ \hline
			\multirow{2}{*}{\textbf{Strategy}} &\multicolumn{1}{c|}{\multirow{2}{*}{
					\textbf{MU--LP} }} & \multicolumn{1}{c|}{\multirow{2}{*}{\textbf{1-layer RS}} }&\multicolumn{1}{c|}{ \multirow{2}{*}{\textbf{Generalized RS}}} & \multicolumn{2}{c|}{\textbf{NOMA}}   \\ \cline{5-6} 
			&                         &                             &                                 &\multicolumn{1}{c|}{ \textbf{SC--SIC per group} }&\multicolumn{1}{c|}{ \textbf{SC--SIC}} \\ \hline
			\textbf{Encoder complexity}       &   Encode  $K+1$ streams                    &          Encode $K+1$ streams                      &        Encode  $2^K-1$ streams                          &      Encode  $K+1$ streams              &     Encode  $K$ streams     \\ \hline
			\textbf{Scheduler complexity}      &        More complex since MU--LP relies on pairing semi-orthogonal users with similar channel gains                &           Simpler to cope with any user deployments without user grouping and ordering issues                 &         Complex to decide upon $\prod_{k=2}^{K-1} \binom{K}{k}! $   decoding orders                      &       Complex to decide upon $\sum_{k=1}^KS(K,k)$ grouping method and  at most $K!$  decoding order  for each grouping method          &  Complex to find aligned users with channel disparity, should decide upon $K!$ decoding orders     \\ \hline
			\textbf{Receiver complexity}      &            1 layer of SIC              &      1 layer of SIC                        &          $2^{K-1}$       layers of SIC                 &      $K-1$ layers of SIC              &  $K-1$ layers of SIC        \\ \hline
	\end{tabular}}
\vspace{-1.5mm}
\end{table*}

 \par In the three-user SC--SIC per group-assisted NOUM transmission, the users are separated into two different groups. Users within each group are served using SC--SIC while the users across the groups are served using MU--LP\cite{mao2017rate}.  As the inter-group interference is mitigated using MU--LP, none of the unicast messages can be encoded with the multicast message $W_0$. One more layer of SIC is required to separate the multicast and unicast streams in the SC--SIC per group-assisted NOUM transmission. Each user first decodes the multicast stream with the highest priority. A three-user example is illustrated in Fig. \ref{fig: scsic per group transmission model}. The users are divided into two different groups with user-1 in group 1 while user-2 and user-3 in group 2.  As there are two users in group 2,  the decoding order is required to be optimized. The decoding order of the unicast messages for the users in group 2 is denoted by $\pi_2$. In  Fig. \ref{fig: scsic per group transmission model}, the decoding order in group 2 is fixed to $\pi_2=2\rightarrow3$. 

 \par Due to the page limitation, the detailed NOMA strategies are not specified. If the readers fully understand Fig. \ref{fig: NOMA transmission} as well as the application of NOMA in the unicast-only transmission discussed in Section 3.2 of \cite{mao2017rate}, the system model of `SC--SIC' and `SC--SIC per group' in the NOUM transmission will be easily traced out.

 \par \textit{Remark 4: Following \cite{mao2017rate}, both the proposed two NOMA-assisted NOUM transmission strategies are sub-strategies of the generalized RS-assisted NOUM. The transmitter complexity of SC--SIC per group-assisted NOUM is higher than the SC--SIC-assisted and the generalized RS-assisted NOUM  since the decoding order and user grouping are required to be optimized together with the precoder.  
A qualitative comparison of the complexity of all the strategies is illustrated in Table \ref{tab: complexity}, where the total number of user grouping methods  to be considered in SC--SIC per group is $\sum_{k=1}^KS(K,k)$. $S(K,k)$ is the number of ways of partitioning a set of $K$ elements into $k$ nonempty sets which is known as a Stirling set number \cite{riordan2012introduction}. It is computed from the sum $S(K,k)=\frac{1}{k!}\sum_{i=0}^k(-1)^i\binom{k}{i}(k-i)^K$. From Table \ref{tab: complexity}, we obtain that 1-layer RS has the simplest scheduler complexity while maintaining the same low encoder and receiver complexity as MU--LP. Since the generalized RS has the highest encoder and receiver complexity while SC--SIC per group has the highest scheduler complexity, both strategies are preferred to be applied to the scenarios when $K$ is small so as to achieve a better tradeoff between the performance improvement and transmitter/receiver complexity. }

\vspace{-1.2mm}
\section{Optimization Frameworks}
\vspace{-0.5mm}
\label{sec: algorithm}
In this section, we specify the optimization frameworks proposed to solve the  WSR   
and EE maximization problems, respectively. 

\vspace{-4mm}
\subsection{WMMSE-based AO algorithm for WSR problems}
\vspace{-0.5mm}
\par  The WMMSE algorithm to solve the sum rate maximization problem in RS without a multicast message is proposed in \cite{RS2016hamdi}.
 It is extended to solve the WSR maximization problems in this work. We firstly explain the procedure to solve the Problem $\textrm{WSR}_{\textrm{1-layer RS}}$ and then specify how the WSR problem of MU--LP, the generalized RS and NOMA can be solved correspondingly.

\par Considering 1-layer RS, user-$k$ decodes the super-common stream $s_{0}$ and the private stream $s_{k}$ sequentially using one  layer of SIC. $s_{0}$ and $s_{k}$ are respectively estimated using the equalizers $g_{k,0}$ and $g_{k}$. Once  $s_{0}$  is successfully decoded by $\hat{s}_{0}=g_{k,0}y_k$ and removed from $y_k$, $s_{k}$ is decoded by $\hat{s}_{k}=g_{k}(y_k-\mathbf{h}_k^H\mathbf{{p}}_{0}\hat{s}_{0})$. The Mean Square Errors (MSEs) of decoding $s_{0}$ and $s_{k}$ are calculated as 
	\vspace{-1mm}
\begin{equation}
\label{eq:MSE}
		\begin{aligned}
		&\varepsilon_{k,0}\triangleq\mathbb{E}\{|\hat{s}_{k,0}-s_{k,0}|^{2}\}=|g_{k,0}|^2T_{k,0}-2\Re\{g_{k,0}\mathbf{h}_k^H\mathbf{p}_{0}\}+1,\\
		&\varepsilon_{k}\triangleq\mathbb{E}\{|\hat{s}_{k}-s_{k}|^{2}\}=|g_{k}|^2T_k-2\Re\{g_{k}\mathbf{h}_k^H\mathbf{p}_k\}+1,
		\end{aligned}
			\vspace{-1mm}
	\end{equation}
	where $T_{k,0}\triangleq|\mathbf{h}_k^H\mathbf{p}_{0}|^2+\sum_{j\in\mathcal{K}}|\mathbf{h}_k^H\mathbf{p}_{j}|^2+1$  and  $T_{k}\triangleq T_{k,0}-|\mathbf{h}_k^H\mathbf{p}_{0}|^2$. By solving $\frac{\partial\varepsilon_{k,0}}{\partial g_{k,0}}=0$ and $\frac{\partial\varepsilon_{k}}{\partial g_{k}}=0$, the optimum MMSE equalizers are given by
	\vspace{-1mm}
	\begin{equation}
	\label{eq:MMSE}
			\begin{aligned}
			&g_{k,0}^{\mathrm{MMSE}}=\mathbf{p}_{0}^H\mathbf{h}_k{T}_{k,0}^{-1},\,\,
			g_{k}^{\mathrm{MMSE}}=\mathbf{p}_k^H\mathbf{h}_k{T}_{k}^{-1}.
			\end{aligned}
	\vspace{-1mm}
		\end{equation}
		 Substituting (\ref{eq:MMSE}) into (\ref{eq:MSE}), the MMSEs become
$\varepsilon_{k,0}^{\textrm{MMSE}} ={({T}_{k,0}-|\mathbf{h}_k^H\mathbf{p}_{k}|^2)}/{{T}_{k,0}}$ and 
$				\varepsilon_{k}^{\textrm{MMSE}} ={(T_{k}-|\mathbf{h}_k^H\mathbf{p}_{k}|^2)}/{T_{k}}.$
Then the SINRs of $s_0$ and $s_k$ can be transformed to $\gamma_{k,0}={1}/{\varepsilon_{k,0}^{\textrm{MMSE}}}-1$ and
			$\gamma_{k}={1}/{\varepsilon_{k}^{\textrm{MMSE}}}-1$.
			The rates become $R_{k,0}=-\log_{2}(\varepsilon_{k,0}^{\textrm{MMSE}})$ and $R_{k}=-\log_{2}(\varepsilon_{k}^{\textrm{MMSE}})$. 

By introducing the positive weights ($w_{k,0},w_{k}$), the WMSEs of decoding $s_0$ and $s_k$ at user-$k$ are defined as
\vspace{-2mm} 
	\begin{equation}
\xi_{k,0}\triangleq w_{k,0}\varepsilon_{k,0}-\log_{2}(w_{k,0}),\,\,\xi_{k}\triangleq w_{k}\varepsilon_{k}-\log_{2}(w_{k}).
\vspace{-1mm} 
	\end{equation}
Then the Rate-WMMSE relationships are established as
\vspace{-1.5mm} 
\begin{equation}
\label{eq: rate-wmmse}
	\begin{aligned}
\xi_{k,0}^{\textrm{MMSE}}&\triangleq\min_{w_{k,0},g_{k,0}}\xi_{k,0}=1-R_{k,0},\\
	\xi_{k}^{\textrm{MMSE}}&\triangleq\min_{w_{k},g_{k}}\xi_{k}=1-R_{k}.
	\end{aligned}
			\vspace{-1.5mm}
\end{equation}
where  $\xi_{k,0}^{\textrm{MMSE}}$ and $\xi_{k}^{\textrm{MMSE}}$ are obtained by substituting the optimum MMSE equalizers  $g_{k,0}^*$, $g_{k}^*$ and the optimum MMSE weights  $w_{k,0}^*$, $w_{k}^*$ back to the WMSEs. 
The optimum MMSE equalizers and  MMSE weights are $g_{k,0}^*=g_{k,0}^{\textrm{MMSE}}$ and $g_{k}^*=g_{k}^{\textrm{MMSE}}$, respectively $w_{k,0}^*=w_{k,0}^{\textrm{MMSE}}\triangleq(\varepsilon_{k,0}^{\textrm{MMSE}})^{-1}$ and $w_{k}^*=w_{k}^{\textrm{MMSE}}\triangleq(\varepsilon_{k}^{\textrm{MMSE}})^{-1}$. They are derived by checking the first order optimality conditions.

							
 \par   Based on the Rate-WMMSE relationships in (\ref{eq: rate-wmmse}), Problem (\ref{eq: rs}) is equivalently transformed   into the WMMSE problem
          \vspace{-1.5mm} 
\begin{subequations}
		\label{eq: rs wmmse}
		\begin{align}
	&\min_{\mathbf{{P}}, \mathbf{x},\mathbf{w},\mathbf{g}} \sum_{k\in\mathcal{K}}u_{k}\xi_{k,tot} \label{object wmmse}\\
		\mbox{s.t.}\quad
		&X_{k,0}+ \xi_{k,0}\leq 1-R_k^{th}, \forall k\in\mathcal{K}\\
		&X_{0}+\sum_{j\in\mathcal{K}}X_{j,0}+1\geq \xi_{k,0}, \forall k\in\mathcal{K}\\
		& X_{0}\leq-R_0^{th}\\
		&  X_{k,0}\leq 0, \forall k\in\mathcal{K}\\
		&	\text{tr}(\mathbf{P}\mathbf{P}^{H})\leq P_{t} 
		\end{align}
	\end{subequations}  
where $\mathbf{x}=[X_{0},X_{1,0},\ldots,X_{K,0}]$ is the transformation of the common rate $\mathbf{c}$. The MMSE weights and equalizers are $\mathbf{w}=[w_{1,0},\ldots,w_{K,0},w_{1},\ldots,w_{K}]$ and $\mathbf{g}=[g_{1,0},\ldots,g_{K,0},g_{1},\ldots,g_{K}]$, respectively $\xi_{k,tot}=X_{k,0}+\xi_{k},\forall k\in\mathcal{K}$.

\par Denote $\mathbf{w}^{\mathrm{MMSE}}$ and $\mathbf{g}^{\mathrm{MMSE}}$ as two vectors formed by the corresponding optimum MMSE equalizers and weights obtained by  minimizing (\ref{object wmmse}) with respect to $\mathbf{w}$ and $\mathbf{g}$, respectively. $(\mathbf{w}^{\mathrm{MMSE}}, \mathbf{g}^{\mathrm{MMSE}})$ satisfies the Karush-Kuhn-Tucker (KKT) conditions of Problem (\ref{eq: rs wmmse}).  Based on  (\ref{eq: rate-wmmse}) and the common rate transformation $\mathbf{c}=-\mathbf{x}$, Problem (\ref{eq: rs wmmse})  can be  transformed to Problem (\ref{eq: rs}). The solution given by ($\mathbf{P}^*,\mathbf{c}^*=-\mathbf{x}^*$) meets the KKT optimality conditions of (\ref{eq: rs}) for any point ($\mathbf{P}^*,\mathbf{x}^*,\mathbf{w}^*,\mathbf{g}^*$) satisfying the KKT optimality conditions of (\ref{eq: rs wmmse}).  Hence,  (\ref{eq: rs}) and  (\ref{eq: rs wmmse}) are equivalent.
Though the joint optimization of ($\mathbf{P},\mathbf{x},\mathbf{w},\mathbf{g}$) in  (\ref{eq: rs wmmse}) is still non-convex, (\ref{eq: rs wmmse}) is convex in each block of $(\mathbf{P},\mathbf{x})$, $\mathbf{w}$, $\mathbf{g}$ by fixing the other two blocks.  The block-wise convexity of (\ref{eq: rs wmmse}) motivates us to use the Alternating Optimization (AO) algorithm to solve the problem. 
Algorithm \ref{WMMSE algorithm} specifies the detailed steps of AO. $(\mathbf{w},\mathbf{g})$ and $(\mathbf{P},\mathbf{x})$ are updated iteratively until the convergence of the WSR. $\mathrm{WSR}^{[n]}$ is the WSR calculated based on the updated $(\mathbf{P},\mathbf{x})$ at iteration $[n]$. The convergence of the AO algorithm is guaranteed \cite{RS2016hamdi} since $\mathrm{WSR}^{[n]}$ is increasing with $n$  and it is bounded above for a given power constraint. Note that the initialization of $\mathbf{P}$ will influence the point of convergence due to the non-convexity of the problem. 
\setlength{\textfloatsep}{5pt}	
\begin{algorithm}[t!]
	\textbf{Initialize}: $n\leftarrow0$, $\mathbf{P}^{[n]}$, $\mathrm{WSR}^{[n]}$\;
	\Repeat{$|\mathrm{WSR}^{[n]}-\mathrm{WSR}^{[n-1]}|\leq \epsilon$}{
		$n\leftarrow n+1$\;
		$\mathbf{P}^{[n-1]}\leftarrow \mathbf{P}$\;
		$\mathbf{w}\leftarrow\mathbf{w}^{\mathrm{MMSE}}(\mathbf{P}^{[n-1]})$; $\mathbf{g}\leftarrow\mathbf{g}^{\mathrm{MMSE}}(\mathbf{P}^{[n-1]})$\;
		update $(\mathbf{x},\mathbf{P})$ by solving (\ref{eq: rs wmmse}) using the updated $\mathbf{w}, \mathbf{g}$;	
	}	
	\caption{WMMSE-based AO algorithm}
	\label{WMMSE algorithm}					
\end{algorithm}
\par When CSIT is imperfect, the sampling-based method proposed in \cite{RS2016hamdi} is adopted to approximate the average rate over the CSIT error distribution for a given channel state estimate. The precoders are designed to maximize the average rate by using the optimization framework described above. The WSR maximization problem of MU--LP, the generalized RS and NOMA are solved by respectively reformulating them into the equivalent WMMSE problem and using the corresponding AO algorithms to solve them.

\vspace{-3mm}
\subsection{SCA-based algorithm for EE problems}	
\vspace{-0.5mm}
\par  The SCA-based algorithm to solve the two-user EE maximization problem of RS without individual QoS rate constraints in the unicast-only transmission is proposed in \cite{mao2018EE}. It is extended to solve the EE maximization problems in the NOUM transmission in this work. We firstly explain the procedure to solve the Problem $\textrm{EE}_{\textrm{1-layer RS}}$ and then specify how the EE problem of MU--LP, the generalized RS and NOMA are solved correspondingly.

\par Comparing with the EE optimization problem (9) in \cite{mao2018EE}, the main difference of  Problem (\ref{eq: EE tot RSMA}) in the NOUM transmission lies in the introduced QoS rate, Constraints (\ref{c1_rs}) and the multicast rate $C_0$ in (\ref{o1_rs}), (\ref{c2_rs}), (\ref{c3_rs}). Similar as \cite{mao2018EE}, we first use scalar variables $\omega^2$,  $z$ and $t$, respectively to represent the WSR, total power consumption and EE metric, then Problem (\ref{eq: EE tot RSMA}) is equivalently transformed into
\vspace{-2mm}
	 	\begin{subequations}
	\label{eq: EE RSMA transform}
	\begin{align}
	\max_{\mathbf{c},\mathbf{{P}},\omega, z,t}&\quad\,\, t\\	
	\mbox{s.t.}\quad	
	&\frac{\omega^{2}}{z}\geq t \label{EE RSMA transform constraint 0}\\ &u_{0}C_{0}+\sum_{k\in\mathcal{K}}u_k\left(C_{k,0}+R_{k}\right)\geq \omega^{2} \label{EE RSMA transform constraint 1}\\	
	&z\geq\frac{1}{\eta}\mathrm{tr}(\mathbf{P}\mathbf{P}^{H})+P_{\textrm{cir}}  \label{EE RSMA transform constraint 2}\\	
	&\textrm{(\ref{c1_rs}) -- (\ref{c5_rs})}
	\end{align}
\end{subequations} 
The equivalence between (\ref{eq: EE RSMA transform}) and (\ref{eq: EE tot RSMA}) is established since (\ref{EE RSMA transform constraint 0})--(\ref{EE RSMA transform constraint 2}) hold with equality at optimum.  By introducing variables  $\bm{\alpha}=[\alpha_1,\ldots,\alpha_K]^T$, Constraints 
(\ref{c1_rs}) and
(\ref{EE RSMA transform constraint 1}) become
\vspace{-3mm}
\begin{subequations}
	\begin{empheq}[left=
	\textrm{(\ref{c1_rs}) , (\ref{EE RSMA transform constraint 1})}\Leftrightarrow\empheqlbrace]{align}
	&C_{k,0}+\alpha_{k}\geq R_{k}^{th}, \forall k \in \mathcal{K} \label{c1_rs transform}\\
	&u_{0}C_{0}+\sum_{k\in\mathcal{K}}u_k\left(C_{k,0}+\alpha_{k}\right)\geq \omega^{2} \label{con: rate non convex}\\
	&R_{k}\geq\alpha_{k}, \forall k \in \mathcal{K} \label{con: rate transform}	
	\end{empheq}
\end{subequations} 
 By adding variables $\bm{\vartheta}=[\vartheta_1,\ldots, \vartheta_K]^T$, Constraint (\ref{con: rate transform}) is transformed into
\vspace{-3mm}
\begin{subequations}
	\begin{empheq}[left=\textrm{(\ref{con: rate transform})}\Leftrightarrow\empheqlbrace]{align}
	&\vartheta_k\geq 2^{\alpha_{k}}, \forall k \in \mathcal{K} \label{con: v and a}\\
	& 1+\gamma_{k}\geq \vartheta_k , \forall k \in \mathcal{K}\label{con: SINR transform}
	\end{empheq}
\end{subequations} 

\par By further introducing $\bm{\beta}=[\beta_1,\ldots,\beta_K]^T$ to represent the interference plus noise at each user to decode its private steam, Constraint (\ref{con: SINR transform})  is transformed into 
\vspace{-2mm}
\begin{subequations}
	\label{con: 1+SINR}
	\begin{empheq}[left=\textrm{(\ref{con: SINR transform})}\Leftrightarrow\empheqlbrace]{align}
	&\frac{\left|\mathbf{{h}}_{k}^{H}\mathbf{{p}}_{k}\right|^{2}}{\beta_k}\geq \vartheta_k-1 , \forall k \in \{1,2\} \label{con: private SINR non-linear}\\
	&\beta_k\geq \sum_{j\neq k}\left|\mathbf{{h}}_{k}^{H}\mathbf{{p}}_{j}\right|^{2}+1 , \forall k \in \{1,2\} \label{con: private noise interference}
	\end{empheq}
\end{subequations}

\par Therefore,  Constraints
	\textrm{(\ref{c1_rs}) and (\ref{EE RSMA transform constraint 1})}  are equivalent to the Constraints
$
\textrm{(\ref{c1_rs transform})}, \textrm{(\ref{con: rate non convex})}, 
\textrm{(\ref{con: v and a})},\textrm{(\ref{con: 1+SINR})}$.
The same method is used to transform Constraint (\ref{c3_rs}). By introducing variable sets $\bm{\alpha}_0=[\alpha_{1,0},\ldots,\alpha_{K,0}]^H$,  $\bm{\vartheta}_0=[\vartheta_{1,0},\ldots, \vartheta_{K,0}]^T$,  $\bm{\beta}_{0}=[\beta_{1,0},\ldots,\beta_{K,0}]^T$, (\ref{c3_rs}) becomes
\vspace{-2mm}
\begin{subequations}
	\label{con: common}
	\begin{empheq}[left=\textrm{(\ref{c3_rs})}\Leftrightarrow\empheqlbrace]{align}
	& C_{0}+\sum_{j\in\mathcal{K}}C_{j,0}\leq \alpha_{k,0},\forall k\in\mathcal{K}\label{con: common rate}\\
	&\vartheta_{k,0}\geq 2^{\alpha_{k,0}} , \forall k \in \mathcal{K} \label{con: common SINR}\\
	&\frac{\left|\mathbf{{h}}_{k}^{H}\mathbf{{p}}_{0}\right|^{2}}{\beta_{k,0}}\geq \vartheta_{k,0}-1 , \forall k \in \mathcal{K}\label{con: common SINR non-linear}\\
	&\beta_{k,0}\geq \sum_{j\in\mathcal{K}}\left|\mathbf{{h}}_{k}^{H}\mathbf{{p}}_{j}\right|^{2}+1 \label{con: common noise interference}
	\end{empheq}
\end{subequations}

\begin{table*}[t!]
	\vspace{-3mm}	
	\centering
	\caption{{Computational complexity comparison of the algorithms using different strategies}}
	\label{tab: computational complexity}
	{
		\begin{tabular}{|l|l|l|l|}
			\hline
			\textbf{Category}                         & \multicolumn{2}{l|}{\textbf{Strategy}}                                        & \textbf{Algorithm 1 (2)}                                                   \\ \hline
			\multirow{2}{*}{\textbf{One-layer SIC}}   & \multicolumn{2}{l|}{\textbf{MU--LP}}                                          & $\mathcal{O}\left([KN_t]^{3.5}\log(\epsilon^{-1})\right)$                                   \\ \cline{2-4} 
			& \multicolumn{2}{l|}{\textbf{1-layer RS}}                                      & $\mathcal{O}\left([KN_t]^{3.5}\log(\epsilon^{-1})\right)$                                   \\ \hline
			\multirow{3}{*}{\textbf{Multi-layer SIC}} & \multicolumn{2}{l|}{\textbf{Generalized RS}}                                  & $\mathcal{O}\left([2^KN_t]^{3.5}\prod_{k=2}^{K-1} \binom{K}{k}!\log(\epsilon^{-1}) \right)$ \\ \cline{2-4} 
			& \multirow{2}{*}{\textbf{NOMA}} & \textbf{SC--SIC per group} & $\mathcal{O}\left(\sum_{k=1}^KS(K,k)[KN_t]^{3.5}\log(\epsilon^{-1})\right)$                 \\ \cline{3-4} 
			&                                                 & \textbf{SC--SIC}            & $\mathcal{O}\left([KN_t]^{3.5}K!\log(\epsilon^{-1})\right)$                                 \\ \hline
	\end{tabular}}
	\vspace{-4mm}	
\end{table*}

\par Hence, Problem (\ref{eq: EE tot RSMA}) is equivalently transformed into
\vspace{-2mm}
\[		
\begin{aligned}
\max_{\substack{\mathbf{c},\mathbf{{P}}, \omega, z,t,\\ \bm{\alpha}_{0},\bm{\alpha},\bm{\vartheta}_{0},\bm{\vartheta},\bm{\beta}_{0},\bm{\beta}}}&\quad\,\, t\\	
\mbox{s.t.}\quad	
&\quad \textrm{(\ref{c2_rs})},\textrm{(\ref{c4_rs})},\textrm{(\ref{c5_rs})},\textrm{(\ref{EE RSMA transform constraint 0})},\textrm{(\ref{EE RSMA transform constraint 2})}  \\
&\quad \textrm{(\ref{c1_rs transform})}, \textrm{(\ref{con: rate non convex})}, 
\textrm{(\ref{con: v and a})},\textrm{(\ref{con: 1+SINR})},	\textrm{(\ref{con: common})} 
\end{aligned}
\]
\par However, Constraints (\ref{EE RSMA transform constraint 0}), (\ref{con: private SINR non-linear}) and (\ref{con: common SINR non-linear}) are non-convex. 
Linear approximation methods adopted in \cite{mao2018EE} are used to approximate the non-convex part of the constraints in each iteration. 
Left side of  (\ref{EE RSMA transform constraint 0}) is approximated at the point ($\omega^{[n]},z^{[n]}$) of the $n$th iteration by
$
\frac{\omega^{2}}{z}\geq\frac{2\omega^{[n]}}{z^{[n]}}\omega-(\frac{\omega^{[n]}}{z^{[n]}})^{2}z\triangleq\Omega^{[n]}(\omega,z)
$.
The left side of  (\ref{con: private SINR non-linear}) is approximated at the point ($\mathbf{p}_k^{[n]},\beta_{k}^{[n]}$) as
$
{\left|\mathbf{{h}}_{k}^{H}\mathbf{{p}}_{k}\right|^{2}}/{\beta_{k}}\geq{2\mathrm{Re}((\mathbf{{p}}_{k}^{[n]})^{H}\mathbf{{h}}_{k}\mathbf{{h}}_{k}^{H}\mathbf{{p}}_{k})}/{\beta_{k}^{[n]}}-({|\mathbf{{h}}_{k}^{H}\mathbf{{p}}_{k}^{[n]}|}/{\beta_{k}^{[n]}})^{2}\beta_{k}\triangleq\Psi_{k}^{[n]}(\mathbf{{p}}_{k},\beta_{k})
$. Similarly, the left side of  (\ref{con: common SINR non-linear})   is approximated at the point  ($\mathbf{p}_0^{[n]},\beta_{k,0}^{[n]}$) by $\Psi_{k,0}^{[n]}(\mathbf{{p}}_{0},\beta_{k,0})={2\mathrm{Re}((\mathbf{{p}}_{0}^{[n]})^{H}\mathbf{{h}}_{k}\mathbf{{h}}_{k}^{H}\mathbf{{p}}_{0})}/{\beta_{k,0}^{[n]}}-({|\mathbf{{h}}_{k}^{H}\mathbf{{p}}_{0}^{[n]}|}/{\beta_{k,0}^{[n]}})^{2}\beta_{k,0}$. Based on the above approximations, Problem (\ref{eq: EE tot RSMA}) is approximated at iteration $n$ as
\vspace{-1mm}
\begin{equation}
\label{eq: final problem}
\begin{aligned}
\max_{\substack{\mathbf{c},\mathbf{{P}}, \omega, z,t,\\ \bm{\alpha}_{0},\bm{\alpha},\bm{\vartheta}_{0},\bm{\vartheta},\bm{\beta}_{0},\bm{\beta}}}&\quad\,\, t\\	
\mbox{s.t.}\quad	
&\quad \Omega^{[n]}(\omega,z)\geq t \\
&\quad \Psi_{k}^{[n]}(\mathbf{{p}}_{k},\beta_{k}) \geq \vartheta_{k}-1, \forall k \in\mathcal{K}\\
&\quad \Psi_{k,0}^{[n]}(\mathbf{{p}}_{0},\beta_{k,0})\geq \vartheta_{k,0}-1, \forall k \in \mathcal{K}\\
&\quad  \textrm{(\ref{c2_rs})},\textrm{(\ref{c4_rs})},\textrm{(\ref{c5_rs})},\textrm{(\ref{c1_rs transform})},\textrm{(\ref{con: rate non convex})},
\textrm{(\ref{con: v and a})}, \\
&\quad\textrm{(\ref{con: private noise interference})},\textrm{(\ref{con: common rate})},\textrm{(\ref{con: common SINR})},\textrm{(\ref{con: common noise interference})}
\end{aligned}
\end{equation}
Problem (\ref{eq: final problem})  is convex and can be solved using CVX in Matlab \cite{grant2008cvx}. The details of the SCA-based algorithm is specified in Algorithm \ref{SCA algorithm}. In each iteration $[n]$, the approximate Problem (\ref{eq: final problem})  defined around the solution of iteration $[n-1]$ is solved.

\begin{algorithm}[h!]	
	\textbf{Initialize}: $n\leftarrow0$, $t^{[n]},\omega^{[n]}, z^{[n]}$, $\mathbf{P}^{[n]}, \bm{\beta}_0^{[n]}, \bm{\beta}^{[n]}$\;
	\Repeat{$|t^{[n]}-t^{[n-1]}|<\epsilon$}{
		$n\leftarrow n+1$\;
		Solve (\ref{eq: final problem}) using $\omega^{[n-1]}$, $z^{[n-1]}$, $\mathbf{P}^{[n-1]}$, $\bm{\beta}_0^{[n-1]}$, $\bm{\beta}^{[n-1]}$ and denote the optimal values as $\omega^*$, $z^*$, $\mathbf{P}^*$, $\bm{\beta}_0^*$, $\bm{\beta}^*$ \;
		Update $t^{[n]}\leftarrow t^*$, $\omega^{[n]}\leftarrow\omega^*$, $z^{[n]}\leftarrow z^*$,   $\mathbf{P}^{[n]}\leftarrow \mathbf{P}^*$, $\bm{\beta}_0^{[n]}\leftarrow\bm{\beta}_0^*$,   $\bm{\beta}^{[n]}\leftarrow\bm{\beta}^*$\;				
	}
	
	\caption{SCA-based beamforming algorithm}
	\label{SCA algorithm}		
\end{algorithm}

\par \textit{Initialization}:
The precoder $\mathbf{P}^{[0]}$ is initialized by finding the feasible beamformer satisfying the constraints (\ref{c1_rs})--(\ref{c5_rs}). We assume in the initialization that  $ C_{0}=R_0$, $C_{k,0}=0,\forall k \in\mathcal{K}$. The non-convex rate constraints are relaxed based on the convex relaxations introduced in \cite{chen2017joint}. After relaxation, the feasibility problem becomes a Second Order Cone Problem (SOCP)  and can be solved by the standard solvers in MatLab. 
$\omega^{[0]}$, $z^{[0]}$, $\beta_k^{[0]}$ and $\beta_{k,0}^{[0]}$ are initialized by respectively replacing the inequalities of  (\ref{EE RSMA transform constraint 1}), (\ref{EE RSMA transform constraint 2}), (\ref{con: private noise interference}) and (\ref{con: common noise interference}) with equalities.

\textit{Convergence Analysis}:  The solution of Problem (\ref{eq: final problem}) in iteration $[n]$ is also a feasible solution of the problem in iteration $[n+1]$ since the approximated Problem (\ref{eq: final problem})  in iteration  $[n+1]$ is defined around the solution of iteration $[n]$. Therefore, the EE objective $t^{[n+1]}$ is larger than or equal to $t^{[n]}$.  Algorithm \ref{SCA algorithm} generates a nondecreasing sequence of objective values. Moreover, the EE objective $t$ is bounded above by the transmit power constraint. Hence, Algorithm \ref{SCA algorithm} is guaranteed to converge while the global optimality of the achieved solution can not be guaranteed.

\par The EE maximization problem of MU--LP, the generalzied RS and NOMA are solved by respectively approximating them using the above transformation and approximation, which are then solved iteratively by the corresponding SCA-based beamforming algorithms as well.
\setlength{\textfloatsep}{5pt}

{
	\vspace{-3mm}
	\subsection{Computational complexity analysis}
\par  The computational complexity of both Algorithm \ref{WMMSE algorithm} and Algorithm \ref{SCA algorithm} for all strategies are illustrated in Table \ref{tab: computational complexity} under the assumption that $N_t\geq K$.

\par	At each iteration of Algorithm \ref{WMMSE algorithm}, the MMSE equalizers and weights $(\mathbf{w},\mathbf{g})$ are updated with complexity $\mathcal{O}(K^2N_t)$ for MU--LP and 1-layer RS-assisted strategies. The complexity of the generalized RS to update the equalizers and weights is $\mathcal{O}(2^KK^2N_t)$. Both SC--SIC and SC--SIC per group strategies require complexity $\mathcal{O}(K^3N_t)$ to update the MMSE equalizers and weights. The precoders and common rate vector $(\mathbf{P},\mathbf{x})$ are then updated by solving the SOCP problem. Each SOCP is solved by using interior-point method with computational complexity  $\mathcal{O}([X]^{3.5})$, where $X$ is the total number of variables in the equivalent SOCP problem \cite{ye1997interior}. For each strategy, the number of variables in the SOCP problem is given by $X_{\textrm{MU--LP}}=KN_t+N_t$, $X_{\textrm{1-layer RS}}=KN_t+N_t+K+1$, $X_{\textrm{SC--SIC}}=KN_t+2$, $X_{\textrm{SC--SIC per group}}=KN_t+N_t$, $X_{\textrm{Generalized RS}}=2^KN_t+2^{K-1}K+1-K$. The total number of iterations required for the convergence is  $\mathcal{O}(\log(\epsilon^{-1}))$, where $\epsilon$ the convergence tolerance of Algorithm 1. As specified in Table \ref{tab: computational complexity}, SC--SIC, SC--SIC per group and the generalized RS have high scheduling complexity since Algorithm \ref{WMMSE algorithm} is required to be repeated for all possible decoding order and user grouping at the scheduler. 
	
\par	At each iteration of Algorithm \ref{SCA algorithm}, the approximated SOCP problem is solved. Though additional variables $\bm{\alpha}_{0}$, $\bm{\alpha}$, $\bm{\vartheta}_{0}$, $\bm{\vartheta}$, $\bm{\beta}_{0}$, $\bm{\beta}$ are introduced for convex relaxation, the main complexity still comes from the precoder design.  Algorithm \ref{SCA algorithm} is also required to be repeated for all possible decoding order and user grouping. Therefore, Algorithm \ref{WMMSE algorithm} and Algorithm \ref{SCA algorithm}  has the same worst-case computational complexity approximation.}

	\vspace{-1mm}
\section{Numerical Results of WSR problem}
\label{sec: WSR simulation}
\par  In this section,  we evaluate the WSR of all the transmission strategies in various user deployments and network loads.  
{Besides the typical  underloaded scenarios appearing in MU-MIMO and massive MIMO,  we also investigate overloaded scenarios.   Overloaded regimes, described as the scenarios where the number of served users exceeds the number of transmitting antennas, are becoming more important due to the growing demands for  ultra-high connectivity \cite{enrico2016bruno,hamdi2017bruno,overloaded2019NOMA}. Applications of overloaded scenarios can also be found in multibeam satellite systems where each beam carries the messages of multiple users, forming a multicast group \cite{satellite2017VJ}, as well as in NOMA \cite{NOMA2013YSaito,overloaded2019NOMA}, and coded caching \cite{Enrico2017cach}.}
\vspace{-3mm}
\subsection{Two-user deployments} 
\label{sec: two-user deployment}
\vspace{-0.2mm}

\par 
When $K=2$, the generalized RS model reduces to the 1-layer RS model. Hence, we use `RS' to represent both strategies. RS is still a more general strategy that encompasses MU--LP and SC--SIC-based NOUM strategies as special cases. 
We compare MU--LP, RS and  SC--SIC-based NOUM strategies. { The OMA transmission is considered as the baseline in which a multicast stream is transmitted for both users while the superimposed unicast stream is only intended for a single user. This user decodes the multicast and unicast streams by using SIC while the other user only decodes the multicast stream. }
The receiver complexities of MU--LP, RS and SC--SIC-assisted strategies are the same  when $K=2$. Only one layer of SIC is required.  

\subsubsection{Perfect CSIT}
\label{sec: twouser perfect WSR}
\begin{figure}[t!]
	\vspace{-3mm}
	\centering
    \includegraphics[width=3.03in]{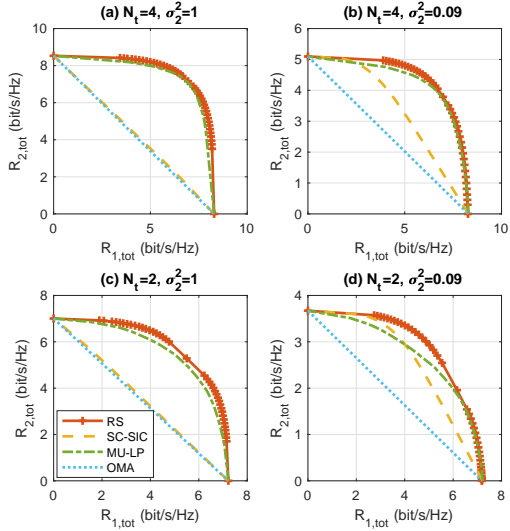}	
	\vspace{-2mm}
	\caption{ Rate region comparison of different strategies in perfect CSIT, averaged over 100 random channel, SNR=20 dB, $\sigma_{1}^2=1$, $R_0^{th}=0.1$  bit/s/Hz.}
	\label{fig: random channel cth01}
	\vspace{-1mm}
\end{figure}

We assume the BS has four or two antennas ($N_t=2, 4$) and serves two single-antenna users.  The initialization of precoders follows the methods used in \cite{mao2017rate,RS2016hamdi}.  SNR is fixed to 20 dB.
The boundary of the rate region is the set of achievable points calculated by solving the WSR maximization problem with various weights assigned to users. The weight of user-1 is fixed to $u_1=1$ for each weight of user-$2$ in $u_2 \in 10^{[-3, -1,-0.95,\cdots ,0.95,1, 3]}$ as used in  \cite{mao2017rate}.  To investigate the largest achievable rate region of the unicast messages, the  rate constraints of the unicast messages are set to 0 in all strategies $R_k^{th}=0,\forall k\in\{1,2\}$.

 We first  consider the channel model when $\mathbf{h}_k$ has independent and identically distributed (i.i.d.) complex Gaussian entries, i.e., $\mathcal{CN}(0,\sigma_{k}^2)$. Fig. \ref{fig: random channel cth01} shows the rate region comparison of different strategies averaged over 100 random channel realizations and $\sigma_{1}^2=1$. When $\sigma_{2}^2=1$ (subfigure (a) and (c)), SC--SIC performs worst as there is no disparity of averaged channel strength. In contrast, MU--LP achieves a rate region close to RS. However, as the number of transmit antenna decreases, the rate region gap between MU--LP and RS becomes more obvious. When $\sigma_{2}^2=0.09$ (subfigure (b) and (d)), the average channel strength disparity between the users is 10 dB. The rate region of SC--SIC comes closer to RS while that of MU--LP becomes worse. RS bridges MU--LP and SC--SIC as well and achieves a better rate region.  In all subfigures, the rate region of OMA is the worst as it is a line segment between the two extremity points of the two users' achievable rate since the unicast rate is dedicatedly allocated to a single user in OMA. The points along the line segment is achieved by time-sharing.  RS exhibits a clear rate region improvement over OMA.

We further investigate specific channel realizations to get some insights into the influence of user angle and channel strength disparity on the system performance. Following the two-user deployment in \cite{mao2017rate},  the channels of the users are realized as $
\mathbf{h}_1=\left[1, 1, 1, 1\right]^H,
\mathbf{h}_2=\gamma\times\left[
1,e^{j\theta},e^{j2\theta},e^{j3\theta}\right]^H.
$
$\gamma$ controls the channel strength difference between the users. $\gamma=1$ and $\gamma=0.3$ represent equal channel strength and  10 dB channel strength difference, respectively. For each  $\gamma$, we consider $\theta\in \left[\frac{\pi }{9},\frac{2\pi }{9},\frac{\pi }{3},\frac{4\pi }{9}\right]$. The user channels are sufficiently aligned when $0<\theta<\frac{\pi }{9}$ while the channels are sufficiently orthogonal when $\frac{4\pi }{9}<\theta<\frac{\pi }{2}$.

\begin{figure}[t!]
	\vspace{-2mm}
	\centering
	\includegraphics[width=3.1in]{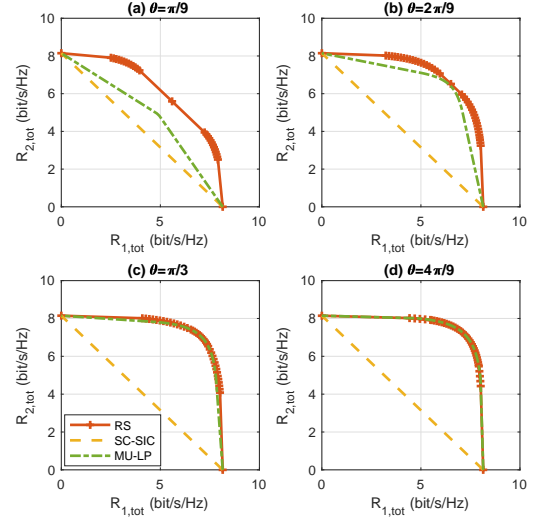}	
	\vspace{-2mm}
	\caption{ Rate region comparison of different strategies in perfect CSIT, $\gamma=1$, $R_0^{th}=0.5$  bit/s/Hz.}
	\label{fig: snr20 bias1 cth05}
	\vspace{-1mm}
\end{figure}
\begin{figure}[t!]
	\vspace{-2mm}
	\centering
	\includegraphics[width=3.2in]{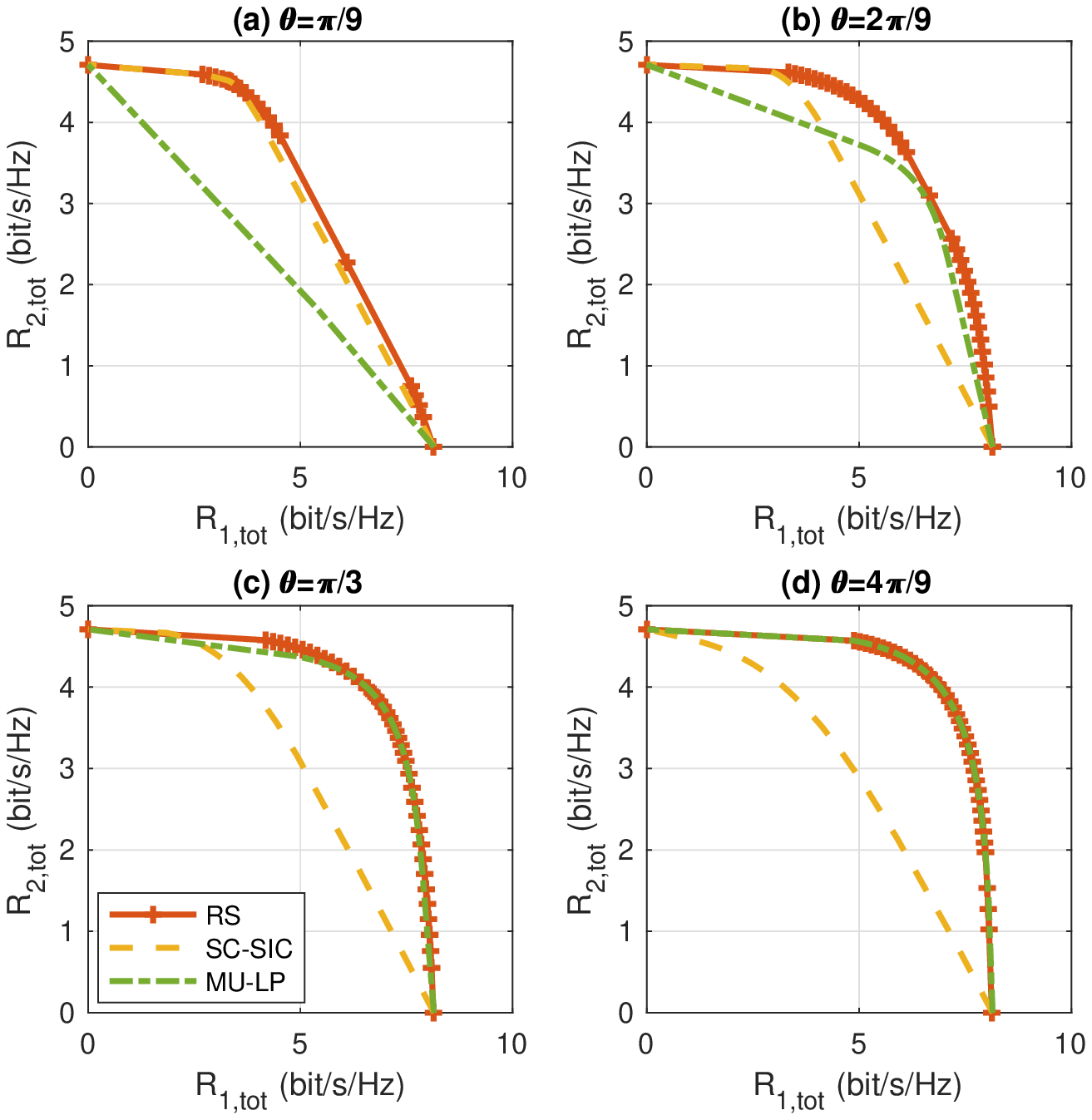}%
	\vspace{-2mm}
	\caption{ Rate region  comparison of different strategies in perfect CSIT, $\gamma=0.3$, $R_0^{th}=0.5$  bit/s/Hz.}
	\label{fig: snr20 bias03 cth05}
	\vspace{-1mm}
\end{figure}
\begin{figure}[t!]
	\vspace{-4mm}
	\centering
\includegraphics[width=3.2in]{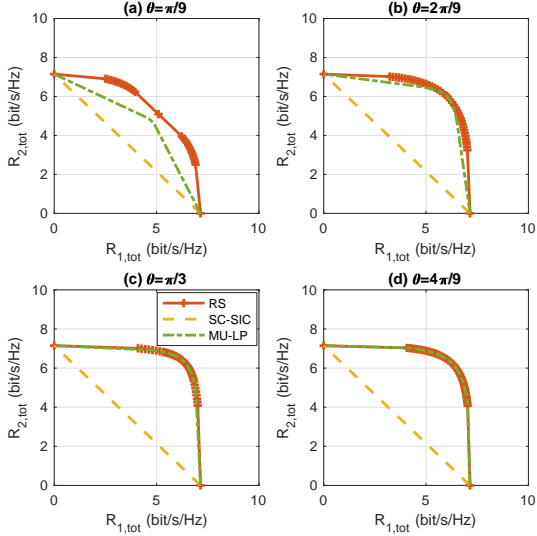}	
	\vspace{-2mm}
	\caption{ Rate region  comparison of different strategies in perfect CSIT, $\gamma=1$, $R_0^{th}=1.5$  bit/s/Hz.}
	\label{fig: snr20 bias1 cth15}
	\vspace{-0mm}
\end{figure}
\par Fig. \ref{fig: snr20 bias1 cth05}--\ref{fig: snr20 bias1 cth15} show the achievable rate region comparison of different strategies in perfect CSIT.
In all figures, the rate region of RS is confirmed to be equal to or larger than that of SC--SIC and MU--LP.  {RS performs well for all investigated channel strength disparities as well as angles between the user channels.}
In contrast, SC--SIC and MU--LP are sensitive to the channel strength disparities and channel angles.
In each figure, RS exhibits a clear rate region improvement over MU--LP when the user channels are closely aligned. 
When the users have similar channel strengths or (semi-)orthogonal channel angles, the performance of SC--SIC is much worse than RS.  Comparing with MU--LP and SC--SIC, RS is more robust to a wide range of channel strength difference and channel angles among users. This WSR gain comes at no additional cost for the receivers since one layer of SIC is required for MU--LP and SC--SIC in the two-user deployments.

\par As the multicast rate constraint $R_0^{th}$ increases, the rate region of each strategy decreases. This can be observed by comparing the corresponding subfigures of Fig. \ref{fig: snr20 bias1 cth05} and Fig. \ref{fig: snr20 bias1 cth15}. However, the rate region gaps among the three strategies decrease when  $R_0^{th}$ increases since a larger portion of the power  is used for transmitting the multicast stream via the super-common stream. RS achieves a better unicast rate region than MU--LP and SC--SIC when a larger portion of the transmit power is allocated to the unicast streams.   Adequate power allocation for the unicast streams allows  RS to  better determine the level of the interference to decode and treat as noise.

\subsubsection{Imperfect CSIT}
\label{sec: WSR two user imperfect CSIT}
When CSIT is imperfect, the estimated channels of user-1 and user-2 are realized as $
\nonumber
\widehat{\mathbf{h}}_{1}=\left[
1,1,1,1\right]^H
$
and 
$
\nonumber
\widehat{\mathbf{h}}_2=\gamma\times\left[
1, e^{j\theta}, e^{j2\theta}, e^{j3\theta}\right]^H$, respectively. The precoders are initialized and designed using the estimated channels $\widehat{\mathbf{h}}_{1},\widehat{\mathbf{h}}_{2}$ and the same methods as stated in \cite{RS2016hamdi,mao2017rate}. The real channel realization  is obtained as $\mathbf{h}_k=\widehat{\mathbf{h}}_{k}+\widetilde{\mathbf{h}}_{k},\forall k\in\{1,2\}$, where $\widetilde{\mathbf{h}}_{k}$ is the estimation error of user-$k$ with independent and identically distributed (i.i.d.) complex Gaussian entries drawn from $\mathcal{CN}(0,\sigma_{e,k}^2)$. The error covariances of user-1 and user-2 are $\sigma_{e,1}^2=P_t^{-0.6}$ and $\sigma_{e,2}^2=\gamma P_t^{-0.6}$, respectively.  
Other unspecified parameters remain consistent with perfect CSIT results.
After generating 1000 different channel error samples for each user,  each point in the rate region is the average rate over the resulting 1000 channels. Note that the average rate is a short-term (instantaneous) measure that captures the expected performance over the CSIT error distribution for a given channel state estimate. {Readers are referred to \cite{RS2016hamdi} for more details of the channel model when CSIT is imperfect. }
\begin{figure}[t!]
	\vspace{-2mm}
	\centering
	\includegraphics[width=3.2in]{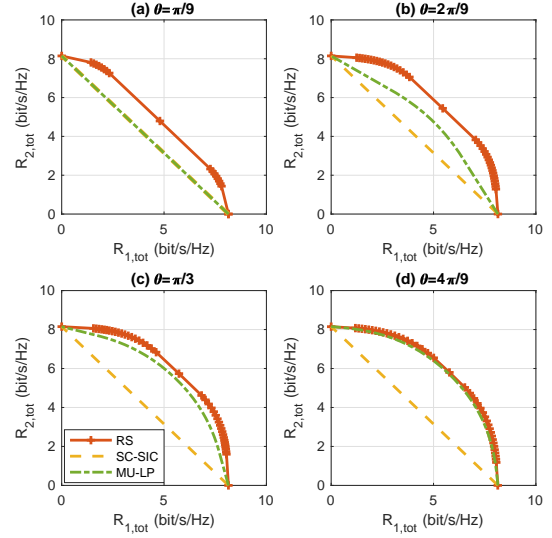}%
	\vspace{-2mm}
	\caption{ Rate region  comparison of different strategies in imperfect CSIT, $\gamma=1$, $R_0^{th}=0.5$  bit/s/Hz.}
	\label{fig: snr20 bias1 cth05 imperfect}
	\vspace{-0mm}
\end{figure}



\par Fig. \ref{fig: snr20 bias1 cth05 imperfect} shows the results when $R_0^{th}=0.5$ bit/s/Hz in imperfect CSIT, $\gamma=1$. Comparing the corresponding figures of perfect and imperfect CSIT (Fig. \ref{fig: snr20 bias1 cth05} and Fig. \ref{fig: snr20 bias1 cth05 imperfect}), we observe that the rate region gap between RS and MU--LP increases in imperfect CSIT. 
RS is more robust to a wide range of CSIT inaccuracy, channel strength difference and channel angles among users. The transmit scheduler of RS is simpler as it copes with any user deployment scenarios.   RS always outperforms MU--LP and SC--SIC.


\vspace{-2mm}
\subsection{Three-user deployments}

In the three-user deployments, 
we compare MU--LP, SC--SIC, SC--SIC per group, 1-layer RS and the generalized RS transmission strategies. In the SC--SIC per group, the grouping method and decoding order are required to be jointly optimized with the precoder in order to maximize the WSR, which results in very high computational burden at the BS as the number of user increases. To reduce the complexity,   we consider a fixed grouping method. We assume user-1 is in group-1 while user-2 and user-3 are in group-2. The decoding order will be optimized together with the precoder. 

\subsubsection{Perfect CSIT} 
\label{sec: threeuser perfect WSR}

\par Following the precoder initialization and channel realizations for three-user deployments in \cite{mao2017rate}, we consider specific channel realizations given by  
$
\mathbf{h}_1=\left[1, 1, 1, 1\right]^H$,
$\mathbf{h}_2=\gamma_1\times[
1, e^{j\theta_1},e^{j2\theta_1}, e^{j3\theta_1}]^H$,
$\mathbf{h}_3=\gamma_2\times[
1, e^{j\theta_2}, e^{j2\theta_2}, e^{j3\theta_2}]^H
$ for the underloaded three-user deployments ($N_t=4$). For the overloaded three-user deployments ($N_t=2$), the channels are realized as $
\mathbf{h}_1=\left[1, 1\right]^H$,
$\mathbf{h}_2=\gamma_1\times[
1, e^{j\theta_1}]^H$,
$\mathbf{h}_3=\gamma_2\times[
1, e^{j\theta_2}]^H.
$
$\gamma_1, \gamma_2$ and  $\theta_1,\theta_2$ are control variables. We assume user-1 and user-2 have equal channel strength ($\gamma_1=1$) and there is a 10 dB channel strength difference between user-1/user-2 and user-3 ($\gamma_2=0.3$). 
For the given set of $\gamma_1,\gamma_2$,  $\theta_1$ adopts value from $\theta_1=\left[\frac{\pi }{9},\frac{2\pi }{9},\frac{\pi }{3},\frac{4\pi }{9}\right]$ and $\theta_2=2\theta_1$. The weights of the users are assumed to be equal to $u_1=u_2=u_3=1$.  The QoS rate requirements of the multicast and unicast messages are assumed to be equal and the rate threshold  is increasing with SNR. For  $\mathrm{SNR}=[0,5,10,15,20,25,30]$ dBs, the corresponding rate constraint vector of message-$j$  is $\mathbf{r}_j^{th}=[0.005, 0.01, 0.05, 0.15, 0.3, 0.4, 0.4]$ bit/s/Hz, $\forall j\in\{0,1,2,3\}$. 
\begin{figure}[t!]
	\vspace{-5.5mm}
    \centering
\includegraphics[width=3.4in]{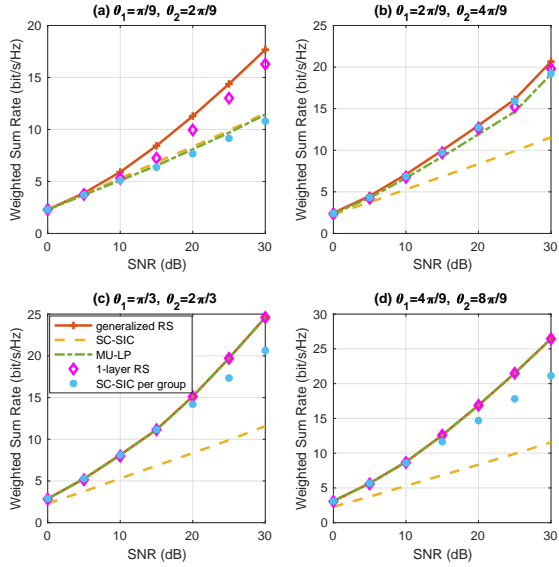}%
	\vspace{-3mm}
	\caption{WSR versus SNR comparison of different strategies for underloaded three-user deployment in perfect CSIT, $\gamma_1=1,\gamma_2=0.3$, $N_t=4$.}
	\label{fig: bias1103 weight111 nt4 perfect}
\end{figure}
\begin{figure}[t!]
	\vspace{-6mm}
	\centering
	\includegraphics[width=3.4in]{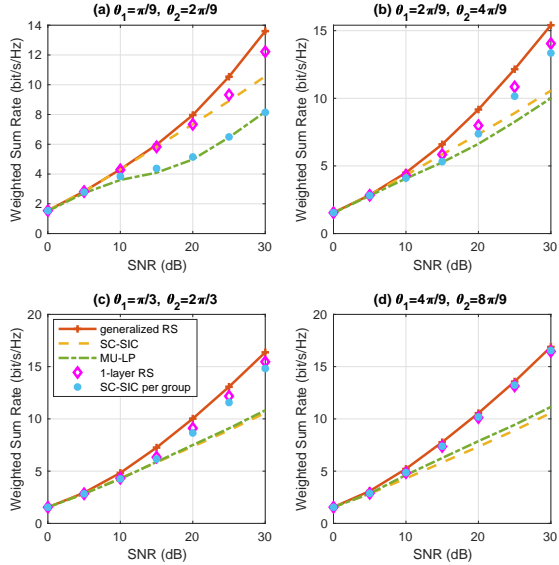}%
	\vspace{-3mm}
	\caption{WSR versus SNR comparison of different strategies for overloaded three-user deployment in perfect CSIT, $\gamma_1=1,\gamma_2=0.3$, $N_t=2$.}
	\label{fig: bias1103 weight111 nt2 perfect}
	\vspace{0mm}
\end{figure}
\par Fig. \ref{fig: bias1103 weight111 nt4 perfect} and Fig. \ref{fig: bias1103 weight111 nt2 perfect} show the results of WSR versus SNR comparison of different strategies in perfect CSIT for the underloaded and overloaded three-user deployments, respectively. RS exhibits a clear WSR gain over 1-layer RS, MU--LP, SC--SIC, SC--SIC per group in both figures. 1-layer RS achieves a more stable performance than MU--LP, SC--SIC, SC--SIC per group as the channel strength disparity and channel angles among users changes. The WSR performance of MU--LP deteriorates as the channel angles among users become smaller (aligned) or the network loads become overloaded. In contrast, the WSR performance of SC--SIC  deteriorates as the channel angles among users become larger or the network load becomes underloaded. SC--SIC per group compensates the shortcomings of SC--SIC. It achieves a better performance than SC--SIC for orthogonal channels or underloaded network loads as it allows the inter-group interference to be treated as noise. Thanks to the ability of partially decoding the interference and partially treating the interference as noise, RS and 1-layer RS are less sensitive to the user channel orthogonality as well as the network loads.
Considering the trade-off between performance and complexity, 1-layer RS is the best choice since it has the lowest receiver complexity and a more robust performance over various user deployments and network loads.

\begin{figure}
	\vspace{-2mm}
\centering
	\begin{subfigure}[b]{0.23\textwidth}		
		\centering
		\includegraphics[width=1.8in]{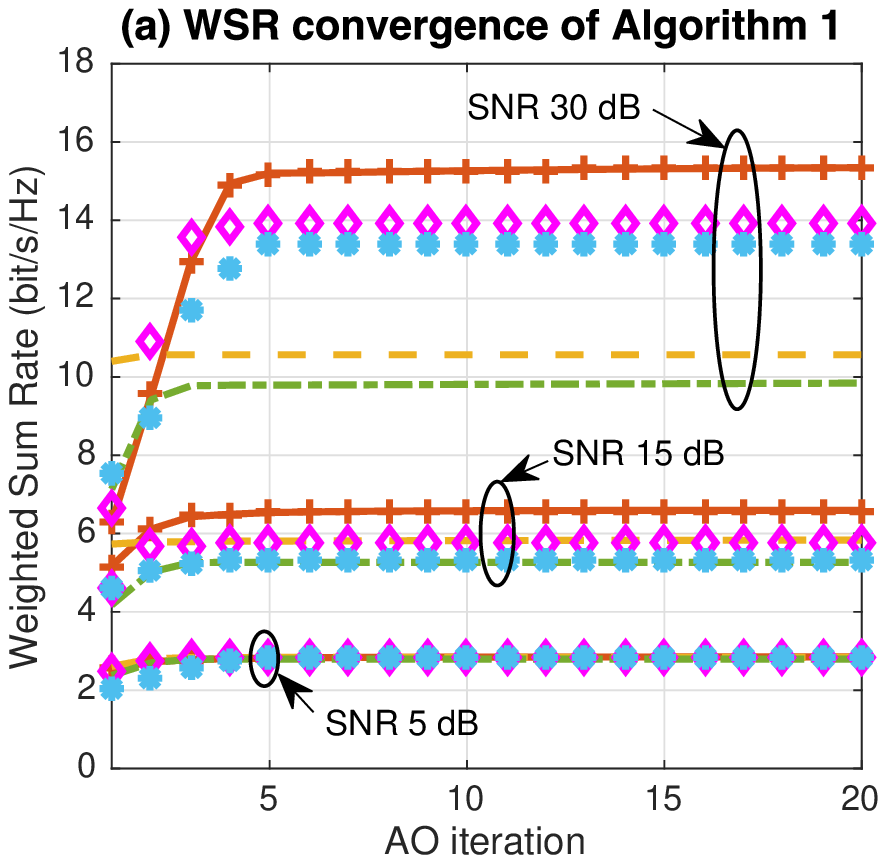}%
	\end{subfigure}%
	~
	\begin{subfigure}[b]{0.23\textwidth}
		\centering
		\includegraphics[width=1.8in]{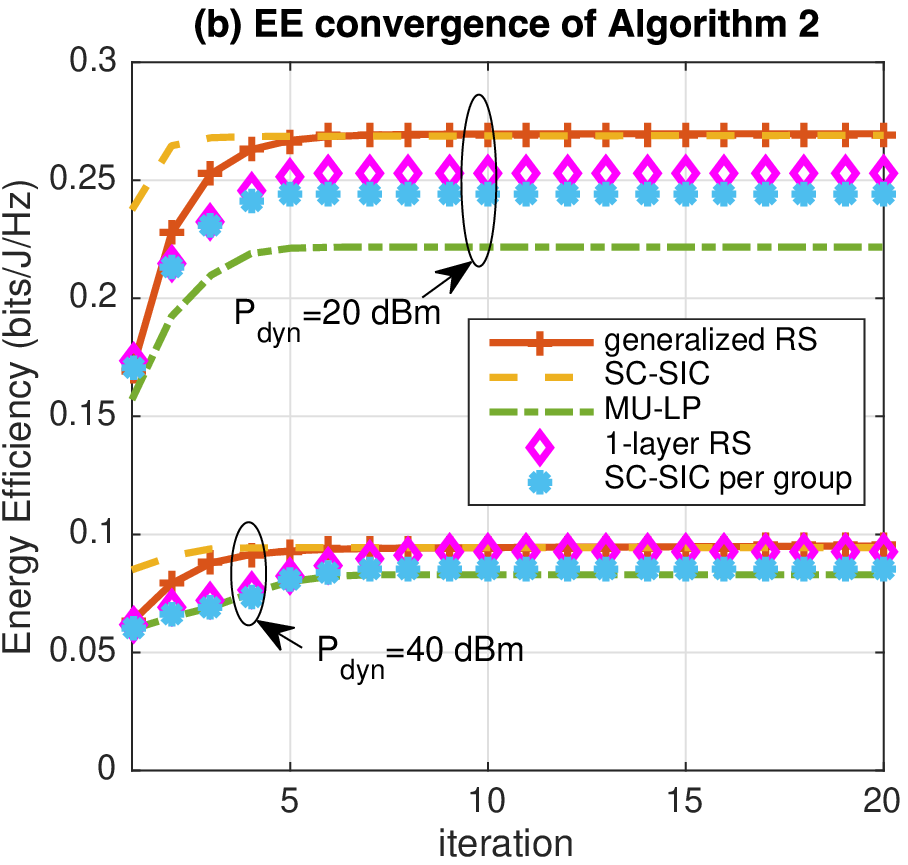}%
	\end{subfigure}%
	\caption{Convergence of the proposed two algorithms with different transmission strategies, $\theta_1=\frac{2\pi }{9}$, $\theta_2=\frac{4\pi }{9}$, $N_t=2$.}
	\vspace{-0mm}
	\label{fig: convergence}
\end{figure}


\par 
 The convergence rates of all the considered transmission strategies for a specific channel realization are analyzed in Fig. \ref{fig: convergence}a. The rate constraints of all messages are equal to the corresponding  value in $\mathbf{r}_j^{th}$ for a given SNR (i.e. when SNR = 5 dB, $R_j^{th}=0.01$ bit/s/Hz, $\forall j\in\{0,1,2,3\}$). As the decoding orders in RS, SC--SIC and SC--SIC per group are required to be optimized with the precoders, the convergence rate of the optimal decoding order that achieves the highest WSR for the corresponding transmission strategy is illustrated in Fig. \ref{fig: convergence}a. For various SNR values, only a few iterations are required for each strategy to converge.  Our proposed WMMSE algorithm solves the WSR problem efficiently.

\begin{figure}
		\vspace{-2mm}
		\centering
		\includegraphics[width=3.4in]{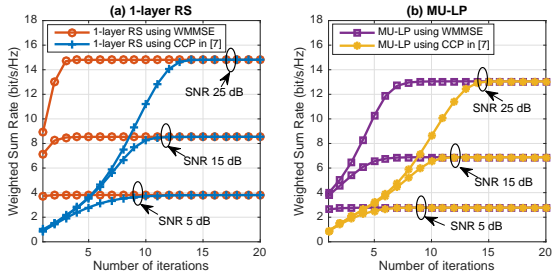}%
		\caption{ Convergence comparison of CCP and WMMSE-based algorithms, $\gamma=1$, $\theta=\frac{2\pi }{9}$, $R_0^{th}=R_k^{th}=0.1$  bit/s/Hz.}
		\vspace{-2mm}
		\label{fig: CCPvsWMMSE}
		\vspace{-1mm}
\end{figure}

The Convex-Concave Procedure (CCP) algorithm proposed in [7] can be adopted to solve the WSR maximization problem by transforming the non-convex SINR constraints into a set of Difference of Convex (DC) constraints and approximated using the first-order Taylor expansion. However, due to the individual QoS rate constraint in the investigated WSR maximization problem, additional variables representing the SINR of users' unicast and multicast streams are introduced, which enlarge the dimension of variables in the SOCP problem to be solved in each iteration.  The convergence speed of using CCP-based algorithm is therefore slower.  
Fig. \ref{fig: CCPvsWMMSE} shows the convergence comparison of CCP and WMMSE-based algorithms using 1-layer RS and MU--LP. For both algorithms, the initialization of precoders $\mathbf{{P}}$ and the channel model are the same as discussed in Section \ref{sec: two-user deployment}.  For the CCP-based algorithm, $\boldsymbol{\rho},\,\boldsymbol{\rho}_{0}$ are initialized by $2^{R_{k}^{th}}-1$ and $2^{R_{0}^{th}}-1$, respectively. We could draw the conclusion that the WMMSE-based algorithm converge faster than the CCP-based algorithm and both algorithms achieve almost the same  performance.

\begin{figure}
			\vspace{-3mm}
	\centering
	\includegraphics[width=2in]{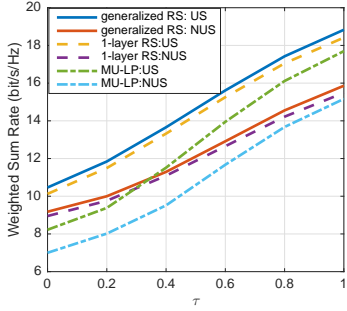}%
	\vspace{-0.5mm}
	\caption{WSR versus CSIT inaccuracy comparison of different strategies over 100 random channel realizations.}
	\label{fig: imperfect CSIT}
\end{figure}

\subsubsection{Imperfect CSIT}


{\par When CSIT is imperfect, we first investigate  random channel realizations. The channel of each user has i.i.d. complex Gaussian entries. 
{Fig. \ref{fig: imperfect CSIT} illustrates the WSR comparison of the generalized RS, 1-layer RS and MU--LP strategies averaged over 100 random channel realizations where $N_t=4$, $u_1=u_2=u_3=1$, $R_0^{th} =R_k^{th}=0.2$ bit/s/Hz and	SNR = 20 dBs.   The  inaccuracy of the channel is controlled by the error covariance  defined as $\sigma_{e,1}^2=\sigma_{e,2}^2=\sigma_{e,3}^2=P_t^{-\tau}$.  $\tau=0$ represents a fixed quality with respect to SNR, e.g. a constant
number of feedback bits, and $\tau=1$ corresponds to perfect CSIT in the DoF sense \cite{RS2016hamdi}. We assume there is a group of $20$ candidate users in the system and only $K=3$ active users are selected. For MU--LP, the User Scheduling (US) algorithm based on channel correlation proposed in  \cite{ZFrateRegion2006}  is adopted. Its worst-case computational complexity is $\mathcal{O}(N_t^3K)$. As RS-based strategies suit to all channel angles, the three users with best channel strength are selected. The computational complexity of such US algorithm is $\mathcal{O}(K)$. 	No User Scheduling (NUS) baseline schemes \textit{{MU--LP: NUS}}, \textit{{1-layer RS: NUS}}, and \textit{{RS: NUS}}  where  users are randomly selected  are illustrated as well. We observe from  Fig. \ref{fig: imperfect CSIT} that the WSR gap between \textit{{1-layer RS: NUS}} (\textit{{1-layer RS: US}}) and \textit{{MU--LP: NUS}} (\textit{{MU--LP: US}}) increases  as $\tau$ decreases. RS is more robust to the inaccuracy of CSIT. Comparing the performance when US  is considered, 1-layer RS outperforms MU--LP but it uses a simpler scheduling algorithm. 
The generalized RS and 1-layer RS without US outperform MU--LP with US when $\tau$ ranges from 0 to 0.3. Therefore, RS-assisted strategies achieves non-negligible gains over MU--LP no matter whether US is considered or not.
	

}
\par When considering specific channel realizations, the precoder initialization and channel realizations follow the methods discussed in the two-user deployment of Section \ref{sec: WSR two user imperfect CSIT}. Readers are also referred to Appendix E in  \cite{mao2017rate} for more details. Other unspecified parameters remain consistent with the perfect CSIT scenarios of Section \ref{sec: threeuser perfect WSR}.
\begin{figure}[t!]
	\vspace{-4.5mm}
	\centering
   \includegraphics[width=3.4in]{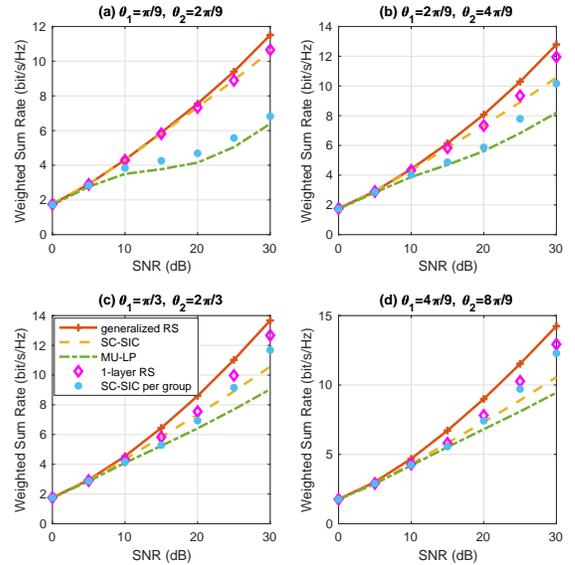}%
	\vspace{-3mm}
	\caption{WSR versus SNR comparison of different strategies for overloaded three-user deployment in imperfect CSIT, $\gamma_1=1,\gamma_2=0.3$, $N_t=2$.}
	\label{fig: bias1103 weight111 nt2 imperfect}
\end{figure}
Fig. \ref{fig: bias1103 weight111 nt2 imperfect} shows the results of WSR versus SNR comparison in the overloaded three-user deployment with imperfect CSIT. Comparing Fig. \ref{fig: bias1103 weight111 nt2 perfect} and Fig. \ref{fig: bias1103 weight111 nt2 imperfect}, the WSR gap between RS and SC--SIC per group/MU--LP is enlarged when the CSIT becomes imperfect.  
Though 1-layer RS has the lowest receiver complexity, it achieves a better WSR than SC--SIC, SC--SIC per group and MU--LP.  


\vspace{-2mm}
\section{Numerical Results of EE problem}
\label{sec: EE simulation}
\par  In this section,  we evaluate the EE performance of all the transmission strategies in various user deployments and network loads. 
\vspace{-3mm}
\subsection{Two-user deployments}
Same as the numerical results of WSR problem, we compare MU--LP, RS  and SC--SIC-assisted NOUM transmission strategies in the two-user deployments. 
\subsubsection{Random channel realizations} 

\label{sec: EE random channel two users}
\par 
We first consider the scenarios when the channel of each user $\mathbf{h}_k$ has i.i.d  complex Gaussian entries with a certain variance, i.e., $\mathcal{CN}(0,\sigma_k^2)$. The variance of entries of $\mathbf{h}_1$ is fixed to 1 ($\sigma_1^2=1$) while the variance of entries of $\mathbf{h}_2$ is varied ($\sigma_2^2=1,0.09$). The BS is equipped with two or four antennas and serves two single-antenna users. Following the simulation parameters used in \cite{mao2018EE}, the static power consumption is $P_{\textrm{sta}}=30$ dBm and the dynamic power consumption is $P_{\textrm{dyn}}=27$ dBm. The power amplifier efficiency is $\eta=0.35$. The weights allocated to the streams are equal to one, i.e., $u_0=u_1=u_2=1$.
 \begin{figure}[t!]
 	\vspace{-3mm}
 	\centering
 	\includegraphics[width=3.3in]{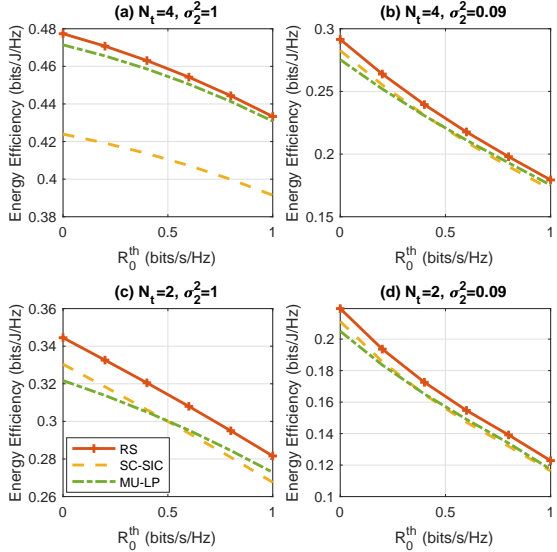}%
 	\vspace{-2mm}
 	\caption{Energy Efficiency versus $R_0^{th}$ comparison of different strategies for two-user deployment in perfect CSIT, averaged over 100 random channels. $R_1^{th}=R_2^{th}=0.5$  bit/s/Hz, SNR = 10 dB.}
 	\label{fig: EE vs R0th}
 	\vspace{0mm}
 \end{figure}

 Fig. \ref{fig: EE vs R0th} shows the results of EE versus the multicast rate requirement $R_0^{th}$ comparison of three transmission strategies for the two-user deployment with perfect CSIT. 
 The proposed RS-assisted NOUM transmission outperforms SC--SIC and MU--LP in all considered user deployments. 
 Comparing subfigure (a) and (c), we observe that the EE gap between RS and MU--LP increases as the number of transmit antenna decreases. MU--LP achieves a better EE performance in the underloaded regime. In contrast, SC--SIC performs better in the overloaded regime. Such observation of the EE performance is consistent with that of the WSR performance.  


\subsubsection{Specific channel realizations}The specific channel realizations and relevant simulation parameters specified in Section \ref{sec: twouser perfect WSR} are considered here. 
In order to investigate the EE region achieved by the unicast streams, the rate allocated to the multicast stream is fixed at $R_0^{th}$, i.e., $C_0=R_0^{th}$. In the following results, we assume $R_0^{th}=0.5$ bit/s/Hz and $u_0=1$.  SNR is fixed at 10 dB and the transmitter is equipped with four tansmit antennas ($N_t=4$). The unspecified parameters remain the same as in the random channel realization section. The EE metric of each unicast stream is defined as the achievable unicast rate divided by the sum power.  The individual EE of user-$k$ is 
$
\textrm{EE}_k={R_{k,tot}}/{(\frac{1}{\eta}\mathrm{tr}(\mathbf{P}\mathbf{P}^{H})+P_{\textrm{cir}})},\forall k\in\{1,2\}.
$
 \begin{figure}[t!]
 	\vspace{-3mm}
 	\centering
 	\includegraphics[width=3.3in]{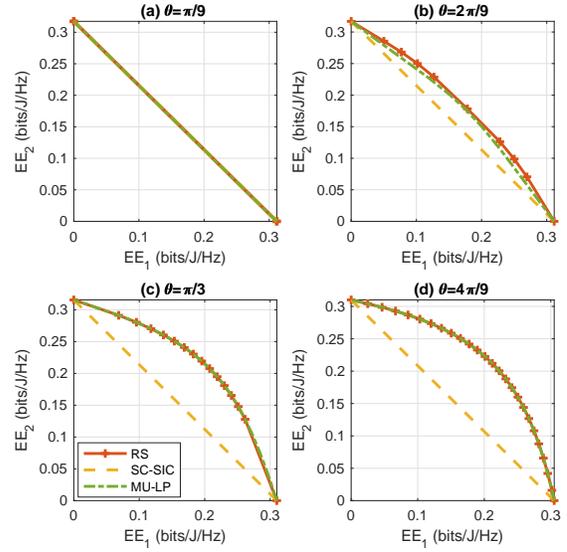}%
 	\vspace{-2mm}
 	\caption{Energy Efficiency region comparison of different strategies for two-user deployment in perfect CSIT, $\gamma=1$.}
 	\label{fig: EE region NT4bias1R0th05}
 	\vspace{0mm}
 \end{figure}

 \begin{figure}[t!]
 	\vspace{-5mm}
 	\centering
 \includegraphics[width=3.3in]{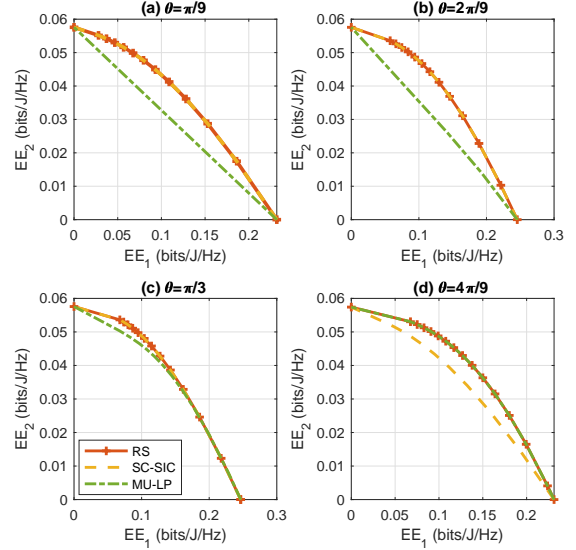}%
 	\vspace{-2mm}
 	\caption{Energy Efficiency region comparison of different strategies for two-user deployment in perfect CSIT, $\gamma=0.3$.}
 	\label{fig: EE region NT4bias03R0th05}
 	\vspace{-0mm}
 \end{figure}
Fig. \ref{fig: EE region NT4bias1R0th05} and Fig. \ref{fig: EE region NT4bias03R0th05} illustrate the EE region of different strategies for the two-user deployment in perfect CSIT, $\gamma=1$ and $\gamma=0.3$, respectively.
The EE region of RS is always larger than or equal to the EE region of MU--LP or SC--SIC in both figures. The EE performance of MU--LP is superior when the user channels are sufficiently aligned. In contrast, the EE performance of SC--SIC is superior when there is a 10 dB channel strength difference or the user channels are aligned.  
Comparing with the EE regions of the unicast-only transmission illustrated in  \cite{mao2018EE}, the EE region improvement of RS in Fig. \ref{fig: EE region NT4bias1R0th05} and Fig. \ref{fig: EE region NT4bias03R0th05}  is not obvious due to the introduced multicast stream. As discussed in Section \ref{sec: EE random channel two users}, the overall optimization space is reduced since part of transmit power is allocated to the multicast stream so as to meet the multicast rate requirement. {Same as the discussion of Fig. \ref{fig: random channel cth01}, the EE region of OMA is a line segment between the two corner points of the users' achievable EE. Therefore, the EE region of OMA is the worst and RS achieves a much better EE region improvement over OMA. }

\begin{figure}[t!]
	\vspace{-2mm}
	\centering	\includegraphics[width=3.3in]{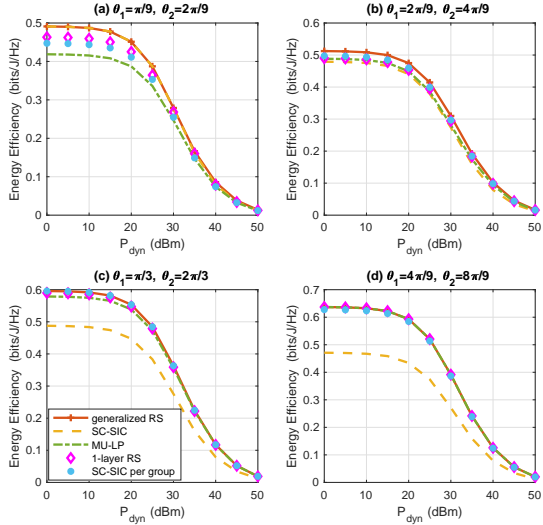}%
	\vspace{-3mm}
	\caption{Energy Efficiency versus $P_{\textrm{dyn}}$ comparison of different strategies for underloaded three-user deployment in perfect CSIT. $N_t=4$.}
	\label{fig: EE dyn NT4cth01}
	\vspace{-1mm}
\end{figure}

\vspace{-3mm}
\subsection{Three-user deployments}
\vspace{-0mm}
In the three-user deployment, we focus on the specific channel realizations and  the influence of different $P_{\textrm{dyn}}$ values on the EE performance is further investigated. Following the three-user WSR analysis, we compare the proposed 1-layer RS, generalized RS, SC--SIC, SC--SIC per group with MU--LP described in previous sections. The specific channel model specified in Section \ref{sec: threeuser perfect WSR} is used in this section. In the following results, the QoS rate constraints of the multicast and  unicast messages are assumed to be equal to 0.1 bit/s/Hz, i.e., $R_0^{th}=R_1^{th}=R_2^{th}=R_3^{th}=0.1$  bit/s/Hz. The weights allocated to the streams are equal to one, i.e., $u_0=u_1=u_2=u_3=1$. SNR is fixed to 10 dB. The channel strength disparities are fixed to $\gamma_1=1,\gamma_2=0.3$.
\begin{figure}[t!]
	\vspace{-4mm}
	\centering
	\includegraphics[width=3.3in]{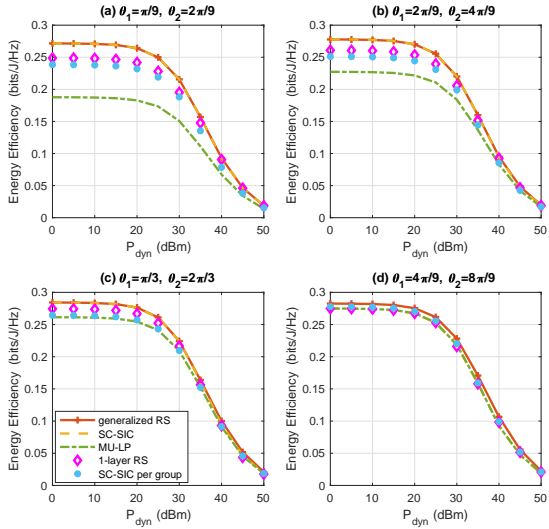}%
	\vspace{-3mm}
	\caption{Energy Efficiency versus $P_{\textrm{dyn}}$ comparison of different strategies for overloaded three-user deployment in perfect CSIT. $N_t=2$.}
	\label{fig: EE dyn NT2cth01}
\end{figure}

\par Fig. \ref{fig: EE dyn NT4cth01} and Fig. \ref{fig: EE dyn NT2cth01} illustrate the EE versus $P_{\textrm{dyn}}$ comparison of different strategies for underloaded and overloaded three-user deployments with perfect CSIT, respectively. In both figures, the generalized RS always outperforms all other strategies.   Though MU--LP and the proposed 1-layer RS have the lowest receiver complexity, the EE performance of  1-layer RS outperforms MU--LP in all figures. It achieves a better EE performance than SC--SIC per group in most simulated user deployments and network loads. 1-layer RS also outperforms SC--SIC when the user channels are sufficiently orthogonal. We conclude that 1-layer RS provides more robust EE performance than MU--LP, SC--SIC and SC--SIC per group towards different user deployments and network loads. 

\par  The EE convergence of all considered transmission strategies for a specific channel realization is analyzed in Fig. \ref{fig: convergence}b.  For various dynamic power values $P_{\textrm{dyn}}$, a few iterations are required for each  strategy to converge. 
Both MU--LP and 1-layer RS-assisted transmission strategies use Algorithm \ref{SCA algorithm} just once to complete the optimization procedure. In contrast,  Algorithm \ref{SCA algorithm} is required to be repeated for each decoding order of RS/SC--SIC/SC--SIC per group-assisted strategies, which results in much higher computational burden at the transmitter especially when  the number of served users is large. The proposed 1-layer RS-assisted NOUM transmission achieves an excellent  tradeoff between EE performance and complexity.


\vspace{-2mm}
\section{Conclusions}
\label{sec: conclusion}
To conclude, we initiate the study of rate-splitting in NOUM transmission by proposing a 1-layer RS and generalized RS-assisted transmission strategies. We also propose two NOMA-assisted transmission strategies, namely, `SC--SIC' and `SC--SIC per group'.  The precoders of all the strategies are designed by maximizing the WSR/EE subject to the sum power constraint and the QoS rate requirements of all  messages. Two low-complexity WMMSE-based and SCA-based optimization frameworks are proposed to solve the WSR and EE maximization problems, respectively.
Numerical results show that the proposed generalized RS-assisted strategy softly bridges and outperforms MU--LP{, OMA} and NOMA in a wide range of user deployments (with a diversity of channel directions, channel strengths and qualities of channel state information at the transmitter) and network loads (underloaded and overloaded regimes). It is a more general and powerful transmission strategy that encompasses MU--LP{, OMA} and NOMA as special cases. 
The proposed 1-layer RS-assisted strategy gets most of the performance benefits of the multi-layer (generalized) RS at a much lower complexity, and is more spectrally efficient and energy efficient than the existing MU--LP-assisted strategy in various user deployments  and network loads. It also achieves a more robust WSR and EE performance than the proposed NOMA-assisted strategies. 
Most importantly,  the high-quality performance of 1-layer RS comes without any increase in the receiver complexity compared with MU--LP and the receiver complexity of 1-layer RS is much lower than the proposed NOMA-based strategies. The one layer SIC in RS is used for the dual purpose of separating the unicast and multicast streams as well as better managing the multi-user unicast interference. Hence, the presence of SIC is better exploited in the proposed 1-layer RS-based strategy. 

\bibliographystyle{IEEEtran}
\vspace{-3mm}
\bibliography{NOUM}	

\begin{thebibliography}{10}
\providecommand{\url}[1]{#1}
\csname url@samestyle\endcsname
\providecommand{\newblock}{\relax}
\providecommand{\bibinfo}[2]{#2}
\providecommand{\BIBentrySTDinterwordspacing}{\spaceskip=0pt\relax}
\providecommand{\BIBentryALTinterwordstretchfactor}{4}
\providecommand{\BIBentryALTinterwordspacing}{\spaceskip=\fontdimen2\font plus
\BIBentryALTinterwordstretchfactor\fontdimen3\font minus
  \fontdimen4\font\relax}
\providecommand{\BIBforeignlanguage}[2]{{%
\expandafter\ifx\csname l@#1\endcsname\relax
\typeout{** WARNING: IEEEtran.bst: No hyphenation pattern has been}%
\typeout{** loaded for the language `#1'. Using the pattern for}%
\typeout{** the default language instead.}%
\else
\language=\csname l@#1\endcsname
\fi
#2}}
\providecommand{\BIBdecl}{\relax}
\BIBdecl

\bibitem{mao2018rate}
Y.~Mao, B.~Clerckx, and V.~O.~K. Li, ``Rate-splitting for multi-antenna
  non-orthogonal unicast and multicast transmission,'' in \emph{Proc. {IEEE}
  Int. Workshop Signal Process. Adv. Wireless Commun. (SPAWC)}, June 2018, pp.
  1--5.

\bibitem{unimulticast2008}
D.~Kim, F.~Khan, C.~V. Rensburg, Z.~Pi, and S.~Yoon, ``Superposition of
  broadcast and unicast in wireless cellular systems,'' \emph{{IEEE} Commun.
  Mag.}, vol.~46, no.~7, pp. 110--117, July 2008.

\bibitem{unimulticast2010second}
U.~Sethakaset and S.~Sun, ``Sum-rate maximization in the simultaneous unicast
  and multicast services with two users,'' in \emph{Proc. IEEE Annu. Symp.
  Pers. Indoor Mobile Radio Commun. (PIMRC)}, Sept 2010, pp. 672--677.

\bibitem{unimulti2012}
Y.~Jia, Z.~Chen, and P.~Ren, ``User selection algorithms for simultaneous
  unicast and multicast services,'' in \emph{Proc. Int. Conf. Wireless Commun.,
  Networking and Mobile Computing}, Sept 2012, pp. 1--4.

\bibitem{Zhao2016LDM}
J.~Zhao, O.~Simeone, D.~Gunduz, and D.~Gomez-Barquero, ``Non-orthogonal unicast
  and broadcast transmission via joint beamforming and {LDM} in cellular
  networks,'' in \emph{Proc. {IEEE} Glob. Commun. Conf. (GLOBECOM)}, Dec 2016,
  pp. 1--6.

\bibitem{Liu2017LDM}
Y.~F. Liu, C.~Lu, M.~Tao, and J.~Wu, ``Joint multicast and unicast beamforming
  for the {MISO} downlink interference channel,'' in \emph{Proc. {IEEE} Int.
  Workshop Signal Process. Adv. Wireless Commun. (SPAWC)}, July 2017, pp. 1--5.

\bibitem{chen2017joint}
E.~Chen, M.~Tao, and Y.~Liu, ``Joint base station clustering and beamforming
  for non-orthogonal multicast and unicast transmission with backhaul
  constraints,'' \emph{{IEEE} Trans. Wireless Commun.}, vol.~17, no.~9, pp.
  6265--6279, Sept 2018.

\bibitem{tervo2017energy}
O.~Tervo, L.~Tran, S.~Chatzinotas, M.~Juntti, and B.~Ottersten,
  ``Energy-efficient joint unicast and multicast beamforming with multi-antenna
  user terminals,'' in \emph{Proc. {IEEE} Int. Workshop Signal Process. Adv.
  Wireless Commun. (SPAWC)}, July 2017, pp. 1--5.

\bibitem{LDM2016}
L.~Zhang, W.~Li, Y.~Wu, X.~Wang, S.~I. Park, H.~M. Kim, J.~Y. Lee, P.~Angueira,
  and J.~Montalban, ``Layered-division-multiplexing: {T}heory and practice,''
  \emph{{IEEE} Trans. Broadcast.}, vol.~62, no.~1, pp. 216--232, March 2016.

\bibitem{3gpp38913}
J.~Krause, ``Study on scenarios and requirements for next generation access
  technology,'' 3GPP TR 38.913, Sept, Tech. Rep., 2016.

\bibitem{ldmtdm2015simeone}
D.~G\'{o}mez-Barquero and O.~Simeone, ``{LDM} versus {FDM}/{TDM} for unequal
  error protection in terrestrial broadcasting systems: {A}n
  information-theoretic view,'' \emph{{IEEE} Trans. Broadcast.}, vol.~61,
  no.~4, pp. 571--579, Dec 2015.

\bibitem{Weing2006Capacity}
H.~Weingarten, Y.~Steinberg, and S.~Shamai, ``On the capacity region of the
  multi-antenna broadcast channel with common messages,'' in \emph{Proc. {IEEE}
  Int. Symp. Inf. Theory (ISIT)}, July 2006, pp. 2195--2199.

\bibitem{Capacity2014Geng}
Y.~Geng and C.~Nair, ``The capacity region of the two-receiver gaussian vector
  broadcast channel with private and common messages,'' \emph{{IEEE} Trans.
  Inf. Theory}, vol.~60, no.~4, pp. 2087--2104, April 2014.

\bibitem{mao2017rate}
Y.~Mao, B.~Clerckx, and V.~O.~K. Li, ``Rate-splitting multiple access for
  downlink communication systems: bridging, generalizing, and outperforming
  {SDMA} and {NOMA},'' \emph{{EURASIP} J. Wireless Commun. Netw.}, vol. 2018,
  no.~1, p. 133, May 2018.

\bibitem{TeHan1981}
T.~Han and K.~Kobayashi, ``A new achievable rate region for the interference
  channel,'' \emph{{IEEE} Trans. Inf. Theory}, vol.~27, no.~1, pp. 49--60, Jan
  1981.

\bibitem{RSintro16bruno}
B.~Clerckx, H.~Joudeh, C.~Hao, M.~Dai, and B.~Rassouli, ``Rate splitting for
  {MIMO} wireless networks: {A} promising {PHY}-layer strategy for {LTE}
  evolution,'' \emph{{IEEE} Commun. Mag.}, vol.~54, no.~5, pp. 98--105, May
  2016.

\bibitem{mao2018EE}
Y.~Mao, B.~Clerckx, and V.~O.~K. Li, ``Energy efficiency of rate-splitting
  multiple access, and performance benefits over {SDMA} and {NOMA},'' in
  \emph{Proc. {IEEE} Int. Symp. Wireless Commun. Syst. (ISWCS)}, Aug 2018, pp.
  1--5.

\bibitem{mao2018networkmimo}
------, ``Rate-splitting multiple access for coordinated multi-point joint
  transmission,'' \emph{Proc. {IEEE} Int. Conf. Commun. (ICC) Workshop}, 2019.

\bibitem{SYang2018SPAWC}
S.~Yang and Z.~Lit, ``A constant-gap result on the multi-antenna broadcast
  channels with linearly precoded rate splitting,'' in \emph{Proc. {IEEE} Int.
  Workshop Signal Process. Adv. Wireless Commun. (SPAWC)}, June 2018, pp. 1--5.

\bibitem{Ahmad2018SPAWC}
A.~A. Ahmad, H.~Dahrouj, A.~Chaaban, A.~Sezgin, and M.~Alouini, ``Interference
  mitigation via rate-splitting in cloud radio access networks,'' in
  \emph{Proc. {IEEE} Int. Workshop Signal Process. Adv. Wireless Commun.
  (SPAWC)}, June 2018, pp. 1--5.

\bibitem{DoF2013SYang}
S.~Yang, M.~Kobayashi, D.~Gesbert, and X.~Yi, ``Degrees of freedom of time
  correlated {MISO} broadcast channel with delayed {CSIT},'' \emph{{IEEE}
  Trans. Inf. Theory}, vol.~59, no.~1, pp. 315--328, Jan 2013.

\bibitem{RS2015bruno}
C.~Hao, Y.~Wu, and B.~Clerckx, ``Rate analysis of two-receiver {MISO} broadcast
  channel with finite rate feedback: {A} rate-splitting approach,''
  \emph{{IEEE} Trans. Commun.}, vol.~63, no.~9, pp. 3232--3246, Sept 2015.

\bibitem{RS2016joudeh}
H.~Joudeh and B.~Clerckx, ``Robust transmission in downlink multiuser {MISO}
  systems: {A} rate-splitting approach,'' \emph{{IEEE} Trans. Signal Process.},
  vol.~64, no.~23, pp. 6227--6242, Dec 2016.

\bibitem{Minbo2016MassiveMIMO}
M.~Dai, B.~Clerckx, D.~Gesbert, and G.~Caire, ``A rate splitting strategy for
  massive {MIMO} with imperfect {CSIT},'' \emph{{IEEE} Trans. Wireless
  Commun.}, vol.~15, no.~7, pp. 4611--4624, July 2016.

\bibitem{RS2016hamdi}
H.~Joudeh and B.~Clerckx, ``Sum-rate maximization for linearly precoded
  downlink multiuser {MISO} systems with partial {CSIT}: {A} rate-splitting
  approach,'' \emph{{IEEE} Trans. Commun.}, vol.~64, no.~11, pp. 4847--4861,
  Nov 2016.

\bibitem{AG2016Gdof}
A.~G. Davoodi and S.~A. Jafar, ``{GDoF} of the {MISO BC}: {B}ridging the gap
  between finite precision {CSIT} and perfect {CSIT},'' in \emph{Proc. {IEEE}
  Int. Symp. Inf. Theory (ISIT)}, July 2016, pp. 1297--1301.

\bibitem{AP2017bruno}
A.~Papazafeiropoulos, B.~Clerckx, and T.~Ratnarajah, ``Rate-splitting to
  mitigate residual transceiver hardware impairments in massive {MIMO}
  systems,'' \emph{{IEEE} Trans. Veh. Technol.}, vol.~66, no.~9, pp.
  8196--8211, Sept 2017.

\bibitem{minbo2017mmWave}
M.~Dai and B.~Clerckx, ``Multiuser millimeter wave beamforming strategies with
  quantized and statistical {CSIT},'' \emph{{IEEE} Trans. Wireless Commun.},
  vol.~16, no.~11, pp. 7025--7038, Nov 2017.

\bibitem{AG2017Gdof}
A.~G. Davoodi and S.~A. Jafar, ``Transmitter cooperation under finite precision
  {CSIT}: {A} {GDoF} perspective,'' \emph{{IEEE} Trans. Inf. Theory}, vol.~63,
  no.~9, pp. 6020--6030, Sept 2017.

\bibitem{enrico2017bruno}
E.~Piovano and B.~Clerckx, ``Optimal {D}o{F} region of the {K}-user {MISO BC}
  with partial {CSIT},'' \emph{{IEEE} Commun. Lett.}, vol.~21, no.~11, pp.
  2368--2371, Nov 2017.

\bibitem{chenxi2017bruno}
C.~Hao, B.~Rassouli, and B.~Clerckx, ``Achievable {D}o{F} regions of {MIMO}
  networks with imperfect {CSIT},'' \emph{{IEEE} Trans. Inf. Theory}, vol.~63,
  no.~10, pp. 6587--6606, Oct 2017.

\bibitem{Lu2018MMSERS}
G.~Lu, L.~Li, H.~Tian, and F.~Qian, ``{MMSE}-based precoding for rate splitting
  systems with finite feedback,'' \emph{{IEEE} Commun. Lett.}, vol.~22, no.~3,
  pp. 642--645, March 2018.

\bibitem{Medra2018SPAWC}
M.~Medra and T.~N. Davidson, ``Robust downlink transmission: {A}n offset-based
  single-rate-splitting approach,'' in \emph{Proc. {IEEE} Int. Workshop Signal
  Process. Adv. Wireless Commun. (SPAWC)}, June 2018, pp. 1--5.

\bibitem{Flores2018ISWCS}
A.~R. Flores, B.~Clerckx, and R.~C. de~Lamare, ``Tomlinson-harashima precoded
  rate-splitting for multiuser multiple-antenna systems,'' in \emph{Proc.
  {IEEE} Int. Symp. Wireless Commun. Syst. (ISWCS)}, Aug 2018, pp. 1--6.

\bibitem{hamdi2017bruno}
H.~Joudeh and B.~Clerckx, ``Rate-splitting for max-min fair multigroup
  multicast beamforming in overloaded systems,'' \emph{{IEEE} Trans. Wireless
  Commun.}, vol.~16, no.~11, pp. 7276--7289, Nov 2017.

\bibitem{Tervo2018SPAWC}
O.~Tervo, L.~Trant, S.~Chatzinotas, B.~Ottersten, and M.~Juntti, ``Multigroup
  multicast beamforming and antenna selection with rate-splitting in multicell
  systems,'' in \emph{Proc. {IEEE} Int. Workshop Signal Process. Adv. Wireless
  Commun. (SPAWC)}, June 2018, pp. 1--5.

\bibitem{hamdi2015multicasting}
H.~Joudeh and B.~Clerckx, ``Sum rate maximization for {MU-MISO} with partial
  {CSIT} using joint multicasting and broadcasting,'' in \emph{Proc. {IEEE}
  Int. Conf. Commun. (ICC)}, June 2015, pp. 4733--4738.

\bibitem{NOMA2013YSaito}
Y.~Saito, Y.~Kishiyama, A.~Benjebbour, T.~Nakamura, A.~Li, and K.~Higuchi,
  ``Non-orthogonal multiple access ({NOMA}) for cellular future radio access,''
  in \emph{Proc. {IEEE} 77th Veh. Technol. Conf. (VTC Spring)}, June 2013, pp.
  1--5.

\bibitem{xu2011improving}
J.~Xu, L.~Qiu, and C.~Yu, ``Improving energy efficiency through multimode
  transmission in the downlink {MIMO} systems,'' \emph{{EURASIP} J. Wireless
  Commun. Netw.}, vol. 2011, no.~1, p. 200, 2011.

\bibitem{riordan2012introduction}
J.~Riordan, \emph{Introduction to combinatorial analysis}.\hskip 1em plus 0.5em
  minus 0.4em\relax Courier Corporation, 2012.

\bibitem{grant2008cvx}
M.~Grant, S.~Boyd, and Y.~Ye, ``{CVX}: {Matlab} software for disciplined convex
  programming,'' 2008.

\bibitem{ye1997interior}
Y.~Ye, \emph{Interior point algorithms: theory and analysis}.\hskip 1em plus
  0.5em minus 0.4em\relax Springer, 1997.

\bibitem{enrico2016bruno}
E.~Piovano, H.~Joudeh, and B.~Clerckx, ``Overloaded multiuser {MISO}
  transmission with imperfect {CSIT},'' in \emph{Proc. 50th Asilomar Conf.
  Signals, Syst. Comput.}, Nov 2016, pp. 34--38.

\bibitem{overloaded2019NOMA}
L.~{Liu}, C.~{Yuen}, Y.~L. {Guan}, Y.~{Li}, and C.~{Huang}, ``Gaussian message
  passing for overloaded massive {MIMO}-{NOMA},'' \emph{{IEEE} Trans. Wireless
  Commun.}, vol.~18, no.~1, pp. 210--226, Jan 2019.

\bibitem{satellite2017VJ}
V.~{Joroughi}, M.~{\'A}. {V{\'a}zquez}, and A.~I. {P{\'e}rez-Neira},
  ``Generalized multicast multibeam precoding for satellite communications,''
  \emph{{IEEE} Trans. Wireless Commun.}, vol.~16, no.~2, pp. 952--966, Feb
  2017.

\bibitem{Enrico2017cach}
E.~{Piovano}, H.~{Joudeh}, and B.~{Clerckx}, ``On coded caching in the
  overloaded miso broadcast channel,'' in \emph{Proc. {IEEE} Int. Symp. Inf.
  Theory (ISIT)}, June 2017, pp. 2795--2799.

\bibitem{ZFrateRegion2006}
T.~Yoo and A.~Goldsmith, ``On the optimality of multiantenna broadcast
  scheduling using zero-forcing beamforming,'' \emph{{IEEE} J. Sel. Areas
  Commun.}, vol.~24, no.~3, pp. 528--541, March 2006.

\end{thebibliography}
\vspace{-10mm}
\begin{IEEEbiographynophoto}{Yijie Mao}
	 is a postdoctoral research associate with the Communications and Signal Processing Group, Department of the Electrical and Electronic Engineering at the Imperial College London. Her research interests include MIMO, rate-splitting and NOMA for 5G and beyond.
\end{IEEEbiographynophoto}
\vspace{-10mm}
\begin{IEEEbiographynophoto}{Bruno Clerckx}
	 is a Reader, the Head of the Wireless Communications and Signal Processing Lab, and the Deputy Head of the Communications and Signal Processing Group, within the Electrical and Electronic Engineering Department, Imperial College London, London, U.K. His area of expertise is communication theory and signal processing for wireless networks.
\end{IEEEbiographynophoto}
\vspace{-10mm}
\begin{IEEEbiographynophoto}{Victor O.K. Li}
	 is Chair of Information Engineering and Cheng Yu-Tung Professor in Sustainable Development at the Department of Electrical and Electronic Engineering at the University of Hong Kong. His research interests include big data, AI, optimization techniques, and interdisciplinary clean energy and environment studies.
\end{IEEEbiographynophoto}

\end{document}